\documentclass[fleqn,usenatbib]{mnras}

\usepackage[T1]{fontenc}
\usepackage{ae,aecompl}

\usepackage{graphicx}	
\usepackage{amsmath}	
\usepackage{amssymb}	
\usepackage{tabularx}

\newcommand{\eagle}{\textsc{eagle}}
\newcommand{\auriga}{\textsc{auriga}}
\newcommand{\Au}{\textsc{Au}}

\newcommand{\kms}{\ensuremath{~\text{km~s}^{-1}}}

\newcommand{\Jr}{\ensuremath{J_{\textnormal{r}}}}
\newcommand{\FJ}{\ensuremath{F\left(\mathbf{J}\right)}}
\newcommand{\FJrL}{\ensuremath{F\left(J_{\textnormal{r}},L\right)}}
\newcommand{\FL}{\ensuremath{F\left(L\right)}}
\newcommand{\FJr}{\ensuremath{F\left(J_{\textnormal{r}}\right)}}
\newcommand{\FE}{\ensuremath{F\left(E\right)}}
\newcommand{\FEL}{\ensuremath{F\left(E,L\right)}}
\newcommand{\FrJrL}{\ensuremath{F\left(r|J_{\textnormal{r}},L\right)}}
\newcommand{\FrEL}{\ensuremath{F\left(r|E,L\right)}}

\newcommand{\dt}{\ensuremath{\mathrm{d}t}}
\newcommand{\dr}{\ensuremath{\mathrm{d}r}}
\newcommand{\Dr}{\ensuremath{\Delta_{\rho}}}

\newcommand{\Da}{\ensuremath{\Delta_{(J_r,L)}}}

\defcitealias{CallinghamMass2019MNRAS.484.5453C}{Callingham19}
\defcitealias{CautunContract}{Cautun19}

\usepackage[normalem]{ulem}

\usepackage{ifthen}
\usepackage{helvet}
\makeatletter
\newcommand{\showfont}{encoding: \f@encoding{},
  family: \f@family{},
  series: \f@series{},
  shape: \f@shape{},
  size: \f@size{}
}
\newcommand{\iffont}[3]{\ifthenelse{\equal{\f@family}{#1}}{#2}{#3}}
\makeatother

\title{The orbital phase space of contracted dark matter halos}

\author[T. M. Callingham et al.]
{\parbox{\textwidth}{
Thomas M. Callingham,$^{1}$\thanks{E-mail: thomas.m.callingham@durham.ac.uk}
Marius Cautun,$^{1,2}$
Alis J. Deason,$^{1}$
Carlos S. Frenk,$^{1}$
Robert J. J. Grand,$^{3}$
Federico Marinacci$^{4}$
and 
Ruediger Pakmor$^{3}$
\vspace{.20cm}} \\
$^{1}$Institute of Computational Cosmology, Department of Physics, University of Durham, South Road, Durham DH1 3LE, UK \\
$^{2}$Leiden Observatory, Leiden University, PO Box 9513, NL-2300 RA Leiden, the Netherlands \\
$^{3}$Max-Planck-Institut f\"{u}r Astrophysik, Karl-Schwarzschild-Str. 1, 85748 Garching, Germany \\
$^4$Department of Physics and Astronomy, University of Bologna, via Gobetti 93/2, 40129 Bologna, Italy \\
\vspace{.10cm}}

\pubyear{2020}

\begin{document}
\label{firstpage}
\pagerange{\pageref{firstpage}--\pageref{lastpage}}
\maketitle

\begin{abstract}
  We study the orbital phase-space of dark matter (DM) halos in the
  \auriga{} suite of cosmological hydrodynamics simulations of Milky
  Way analogues. We characterise halos by their spherical action
  distribution, \FJrL{}, a function of the specific angular momentum,
  $L$, and the radial action, $J_r$, of the DM particles. By comparing
  DM-only and hydrodynamical simulations of the same halos, we
  investigate the contraction of DM halos caused by the accumulation
  of baryons at the centre. We find a small systematic suppression of
  the radial action in the DM halos of the hydrodynamical simulations,
  suggesting that the commonly used adiabatic contraction
  approximation can result in an underestimate of the density
  by $\sim8\%$. We apply an iterative algorithm to contract the
  \auriga{} DM halos given a baryon density profile and halo mass,
  recovering the true contracted DM profiles with an accuracy of
  $\sim15\%$, that reflects halo-to-halo variation. Using this
  algorithm, we infer the total mass profile of the Milky Way's
  contracted DM halo. We derive updated values for the key
  astrophysical inputs to DM direct detection experiments: the DM
  density and velocity distribution in the Solar neighbourhood.
\end{abstract}




\section{Introduction}

The past three decades have seen tremendous advances in our
understanding of galaxies and the dark matter (DM) halos in which
they form. From a theoretical perspective, much effort has been
directed at understanding structure formation in collisionless N-body
simulations, in which both DM and baryons are modelled as a single
dissipationless fluid \citep[see e.g.][for a recent
review]{ZavalaFrenk_DMReview2019arXiv190711775Z}. These are often
referred to as `dark matter-only' (hereafter DMO) simulations. Such
cosmological simulations show that overdense regions first collapse to
form small halos, with larger structures forming hierarchically
through mergers of smaller objects and accretion of diffuse mass
\citep{Frenk1988}. The resulting DM halos have universal density
profiles that are well fit by the Navarro, Frenk \& White (NFW)
form \citep{Navarro1996,Navarro1997}:
\begin{equation} \label{eq:NFWPotDens}
\rho\left(r\right)=\frac{\rho_{s}}{\frac{r}{r_{s}}\left(1+\frac{r}{r_{s}}\right)^{2}}
\;,
\end{equation}
which is characterised by two free parameters: the scale radius,
$r_{\textnormal{s}}$, and the characteristic density, $\rho_{s}$.  The
origins of this simple profile are still debated, with suggestions
including a close connection to the halo merger history or an
attractor solution to entropy driven relaxation
\citep[e.g.][]{Ludlow14_MassConc_2014MNRAS.441..378L,PontzenEntropy2013MNRAS.430..121P}.

This conformity of halos in DMO simulations is broken when baryonic
physics are included in fully hydrodynamical simulations (hereafter
`Hydro'). Such simulations include many of the physical processes
thought to be important in the formation of galaxies, such as gas
cooling and heating, stellar winds, chemical evolution and supernova
and AGN feedback \citep[e.g. see][]{Somerville2015}; they thus have a
much more complex and rich behaviour than their DMO counterparts. In
particular, gas cools and condenses at the halo centre, where it forms
stars.  This results in DM halos that have higher central densities
than a NFW profile, and that are often referred to as having been 
``contracted''.  The amount of DM contraction depends on many factors
including the mass of the central galaxy, its assembly history and 
the orbital distribution of DM particles
\citep[e.g.][]{Gnedin2004,Abadi_ContractShape2010MNRAS.407..435A,Duffy10_ImpactBaryonsDM_2010MNRAS.405.2161D,Schaller2016,Dutton2016,Artale2019,BarnesWhite_1984MNRAS.211..753B,Blumenthal1986ApJ...301...27B}.

DM halos cannot be observed directly, of course, but some of the
properties of the MW halo can be inferred from observations of tracers
of the gravitational potential. The latest \textit{Gaia} data release
(DR2) \citep{GaiaDR22018arXiv180409365G} provides a remarkable
database of full 6D phase-space measurements of stars in the inner
regions of the MW. Combined with other datasets, such as SDSS
\citep{SloanDR142018ApJS..235...42A} and APOGEE
\citep{APOGEE2017AJ....154...94M}, the \textit{Gaia} data have been
used to place tight constraints on the MW's circular velocity curve
\citep{Eilers19Vcirc_2019ApJ...871..120E} and local escape velocity
\citep[e.g.]{Deason19Vesc_2019MNRAS.485.3514D}, and thus have helped
constrain the total mass distribution of the MW. The simplest models
of the MW assume that the DM halo can be described as an NFW profile.
Far from the Galactic Centre, this is a reasonable assumption for the
total mass profile \citep[][hereafter,
\citetalias{CallinghamMass2019MNRAS.484.5453C}]{CallinghamMass2019MNRAS.484.5453C}.
However, to model the inner regions of our galaxy it is essential to
include the mass distributions of its baryonic components such as the
thin and thick disks, the bulge and the stellar halo
\citep[e.g.][]{McMillan2011,McMillan2017}. Previous studies
\citep[e.g.][]{Deason12_BrokenDegenMNRAS.424L..44D, McMillan2017} have
typically found a high halo concentration ($\sim11-12$), which is
unusually large compared to the predictions for MW sized halos from
cosmological simulations (typically $\sim8$ in the \eagle{}
cosmological simulation; \citealt{Schaller2016}). This could be a symptom
of the neglect of the contraction of the DM halo and underlines the
importance of properly accounting for the changes in the DM
distribution induced by the baryonic distribution
\citep[e.g. see][]{CautunContract}.

Several methods have been developed to predict the contracted DM halo
profile in the presence of baryons.  The simplest are different
versions of the adiabatic contraction approximation which assumes that
particle orbits are adiabatic invariants \citep{Eggen1962,Barnes1984}.
An early example of this approach
\cite{Blumenthal1986ApJ...301...27B} effectively assumes that all
particles are on circular orbits, a rather crude approximation that
leads to excessive compression of the orbits.  This method was
improved by \cite{Gnedin2004,Gnedin2011arXiv1108.5736G},
who modified it to take into account that DM particles
are typically on non-circular orbits. However, these improved versions
neglect the fact that DM particles have a distribution of orbits.
\citealt[][(hereafter, \citetalias{CautunContract})]{CautunContract} have
studied the contraction of DM density profiles in the \eagle{} and
\auriga{} simulations and derived an analytic prescription for the
average halo contraction; their approach is unbiased and recovers the
profiles of DM halos in hydro simulations with an accuracy of
$\sim 10\%$ that reflects the halo-to-halo scatter.

While these methods are easy to apply, they neglect important
information and provide only limited understanding.  To model the
effects of contraction properly it is necessary to consider the
complex dynamics within the DM halo. While often viewed as static
profiles, halos are made up of particles moving on
various orbits \citep{ZhuOrbit2017MNRAS.466.3876Z} that conspire to give a steady density profile . For a
halo in equilibrium it follows from the Jeans theorem that the
distribution of the DM particles is solely dependent on integrals of
motion (IoM), with no dependence on phase. The halo can therefore be
described as a collection of orbits defined by IoM instead of
particles. The natural choice for this description are the action
integrals $\left[J_{i}\right]_{i=1,2,3}$. One significant advantage
that the actions have over other IoM is that they are adiabatic invariants,
and thus largely unchanged by sufficiently slow changes in the potential
\citep{BinneyTremaine_1987gady.book.....B}.

The distribution function (DF) of DM particle actions, \FJ{}, can be
thought of as an orbital blueprint of  DM halos that may be use
to calculate various halo properties, such as the density and velocity
anisotropy profiles. If the growth of the baryonic component is a
slow, adiabatic process, then the DM halo is described by the same
\FJ{} as in the absence of baryons, i.e. as in DMO simulations.  Given
this adiabatic assumption, the differences between halos in DMO and
Hydro simulations is induced solely by the deeper gravitational
potential of the baryons which are more centrally concentrated in the
Hydro than in the DMO simulations. While the halos are composed of DM
particles on orbits with the same $\bf{J}$ values, the deeper
potential compresses the DM orbits to lower radii in physical space,
resulting in a higher central density in the Hydro simulations.

The extent to which the adiabatic assumption holds is unclear and
depends on the timescale on which the baryons cool and accumulate at
the centre. If the cooling timescale is shorter than the free-fall
timescale, then the gas undergoes rapid cooling, a non-adiabatic
process. Alternatively, if the cooling timescale is much larger than
the free-fall timescale, the growth of the baryonic component is
adiabatic. There is evidence from analytic arguments
\citep{White_Frenk91} and simulations
\citep[e.g.][]{Correa18Eagle_2018MNRAS.473..538C} that the MW mass
halos are in the slow cooling regime. Once the baryons have settled in
the centre of the halo in a quasi-hydrostatic state they dominate the
central gravitational potential. Subsequent violent events, such as
gas blowouts, can change the inner mass profile rapidly over short
timescales, transferring energy to DM particles in the central region
of halos. For halos that host dwarf galaxies, this process could form
cores in their DM distribution
\citep[e.g.][]{Navarro1996,PontzenCore2012MNRAS.421.3464P,LLamayCores2018arXiv181004186B,BurgerCore2019MNRAS.485.1008B}.

To perform action angle modelling it is necessary to chose a specific
DM action distribution function, \FJ. Typically and, in particular,
for isolated DMO halos, the DF is derived analytically, often assuming
that the DM particle orbits have an isotropic velocity
distribution. Under the adiabatic assumption, these orbits can then be
combined with a given baryon potential to construct a contracted DM
halo. This approach was tested by
\cite{SellwoodContract_2005ApJ...634...70S} against N-body simulations
that included a slowly grown analytic baryonic component. By using
simple action DFs, \citeauthor{SellwoodContract_2005ApJ...634...70S}
found that radially biased halos resist compression while isotropic
distributions end up more compressed (in agreement with the results of
\citealt{Gnedin2004}). In the past decade there have
been significant technical advances in the numerical calculation of
action angles and in the overall modelling framework
\citep{AGAMA2019MNRAS.482.1525V}. More complex action DFs, including one that
produces an approximate NFW density profile in isolation, were
analytically derived by \cite{Posti15DFs_2015MNRAS.447.3060P} and used
in a series of papers of increasing complexity, in which the MW is
modelled with multiple baryon components
\citep{Piffl15BringHalo_2015MNRAS.451..639P,Binney15DFDMHalo_2015MNRAS.454.3653B}. In
the most recent study, by \cite{ColeBinneyCore2017MNRAS.465..798C},
the DF of \citeauthor{Posti15DFs_2015MNRAS.447.3060P} was modified
assuming a non-adibiatic, baryon driven upscattering of low action
orbits, generating a cored DM profile.  

Action angle modelling of halos is frustrated by the lack of a
standard NFW action distribution; currently there is no well
established \FJ{} model that has been rigorously tested in
cosmological simulations. The scatter in DM halo properties, such as
concentration and velocity anisotropy
\citep{Navarro10Diverse_2010MNRAS.402...21N}, adds further complexity
to the task of paramaterising a general action DF of a DM halo. This
scatter likely causes halos described by different DFs to undergo different
amounts of contraction for a given baryonic profile; it is therefore
important to capture the variation with an accurate and flexible
parametrisation of the DF. An alternative approach is to use DFs that
are directly measured in simulations, especially given the recent
increase in the resolution and number of zoom-in simulations of MW
mass halos
\citep[e.g.][]{Fattahi2016,Sawala2016, Auriga_2017MNRAS.467..179G,Garrison-Kimmel2019}.
 
In this paper, we determine the distribution function, \FJ{}, of DM
halos from the \auriga{} simulation suite.  This allows us to infer
accurate DM DFs and, at the same time, sample the breadth of
halo-to-halo scatter in cosmologically representative samples of
MW-mass halos. Each simulation volume has a DMO and a Hydro
simulation. By comparing the halos in one to their counterparts in the
other, we can investigate the validity of the {\em ansatz} that the
formation of MW-like galaxies is an adiabatic process. To do so, we
first discuss how a halo's density and velocity profiles can be
inferred from the action DF, and then test if the halo in the Hydro
simulation (hereafter, Hydro halo) can be recovered by adiabatically
contracting the DF measured in the corresponding DMO simulation
(hereafter, DMO halo).

We illustrate the usefulness of modelling DM halos with an action DF
by a few applications targeted at our Galaxy.  Our approach has
implications beyond the mass profile since it provides accurate
predictions for the DM velocity distribution and its moments. Since we
use the observed baryonic component of the MW, these predictions are
specific to our galaxy and unmatched by conventional approaches. We
illustrate this by predicting the density and velocity distribution
function (VDF) of DM particles in the Solar neighbourhood, key inputs
for direct DM detection experiments
\citep{GreenVelDist2010JCAP...10..034G,GreenLocalDM2017JPhG...44h4001G}.
In the literature, the VDF is usually given by the standard halo model
(SHM), a isothermal DM mass distribution with a Gaussian VDF; however,
high-resolution N-body simulations indicate a somewhat different VDF
\citep{Vogelsberger09LocalDM_2009MNRAS.395..797V}.  In principle,
there is a variety of possible DM DFs, which, in turn, would result in
a variety of VDFs at the Solar neighbourhood
\citep[e.g.][]{MaoVDF_2013ApJ...764...35M}.  The sizeable sample of
halo DFs that we can measure in the \auriga{} simulation suite allows
us to characterise the dispersion in the predicted VDF at the Sun's
location, and thus quantify some of the uncertainties in direct DM
detection experiments.

The structure of the paper is as follows. In Section \ref{Sec:Sims} we introduce
our sample of halos and compare physical profiles and orbital
distributions in the DMO and Hydro cases. In Section~\ref{Sec:HaloConstruction}, we construct
individual orbits, investigate the effects of compression and develop
an iterative method for constructing and contracting physical
halos. We apply this to halos in our sample and study the effects of
adiabatic contraction in general. In Section~\ref{Sec:MW} we contract our halo
sample according to the MW baryon distribution and present our
main results, including predictions for the properties of the MW's local
DM distribution. Finally, in Section \ref{Sec:Conclusion} we summarise our main
conclusions.


\section{ Simulated halos}
\label{Sec:Sims} 

In this paper we use a sample of halos from the \auriga{} project, a
suite of 30 high-resolution cosmological zoom-in simulations of individual MW-like
halos \citep{Auriga_2017MNRAS.467..179G} with halo masses between $1-2\times 10^{12} M_{\odot}$. The halos were selected from
the $100^{3} \, \textnormal{Mpc}^{3}$ periodic cube of the \eagle{}
project, a $\Lambda$CDM cosmological hydrodynamical simulation
\citep{EAGLE2015MNRAS.446..521S}. Using the N-body and moving mesh
magnetohydronymic (MHD) \textsc{arepo} code
\citep{SpringelAREPO2011IAUS..270..203S}, these halos were resimulated
to produce both a dark-matter-only and a full hydrodynamic (hereafter
referred to as DMO and Hydro respectively) zoom-in simulation of each
halo. We primarily use the level 4 resolution sample of $30$ halos,
which we label as Au1 to Au30.
The halos in the Hydro simulations have a DM particle mass of
$\sim 3\times 10^{5} M_{\odot}$ and an initial gas resolution element
of mass $\sim 5\times 10^{4} M_{\odot}$.  For the DMO simulations, the
particle mass is $\sim 3.5\times 10^{5} M_{\odot}$. Both the DMO and
Hydro simulations assume the Planck1 \citep{planck2014} cosmological parameters.

In our analysis we treat the halos as being in near spherical
equilibrium. In reality, no halo perfectly satisfies this criterium
and halos are often out of equilibrium after following minor or major
mergers, before relaxing to equilibrium. To characterise the dynamical
state of a halo we employ the \citet{Neto2007} criteria according to
which a halo is relaxed if:
\begin{enumerate}
    \item The total mass of substructure within $R_{200}$ is less
      than 10 per cent of the total halo mass, $M_{200}$. 
    \item  The distance between the centre of mass and the centre of
      potential of the halo is less than $0.07R_{200}$. 
    \item The virial ratio $2T/\left|U\right| <1.35$, where $T$ is the
      total kinetic energy and $U$ the gravitational potential energy
      of DM particles within $R_{200}$. 
\end{enumerate}

These criteria identify 13 out of the 30 \auriga{} halos as unrelaxed
in either the Hydro or the DMO simulations.  These halos are included
in our sample in order to investigate the dependence and sensitivity
of our analysis to departures from equilibrium. Typically, the halos
relax from the inside out, and the halo outskirts (approximately
around and beyond $R_{200}$) are the least virialised and phase mixed
regions.
We have checked that most of the relaxed \auriga{} halos are
reasonably spherical, especially in the inner regions. For example,
the DM particles within $R_{200}/2$ are characterised by the moment of
inertia with minor-to-major axes ratio, $c/a$, of
$0.76^{+0.08}_{-0.03}$ in the DMO simulations and $0.87^{+0.03}_{-0.06}$ in
the Hydro simulations. The presence of baryons in the Hydro simulations
systematically leads to the formation of more spherical halos, as
shown by earlier studies
\citep[e.g][]{Abadi_ContractShape2010MNRAS.407..435A,Prada2019MNRAS.490.4877P,Zhu2016MNRAS.458.1559Z}. Throughout this
work we have checked that there are no systematic trends that
correlate with the degree of halo asphericity, which suggests that our
spherical dynamics treatment represents a reasonable approximation. 

While not explicitly shown, we have performed the same analysis on the
six \auriga{} halos that were simulated at 8 times better mass
resolution than the level 4 simulations considered. While the baryon
profiles can differ due to the dependence of subgrid physics on
resolution and due to stochastic effects, we find the same results as
for the level 4 simulations. As such, we have chosen to show the
results obtained using the larger level 4 simulation sample to better characterise the halo-to-halo variability.

\subsection{Halo Properties}
\label{SubSec:HaloProp} 

\begin{figure}
	\includegraphics[width=\columnwidth]{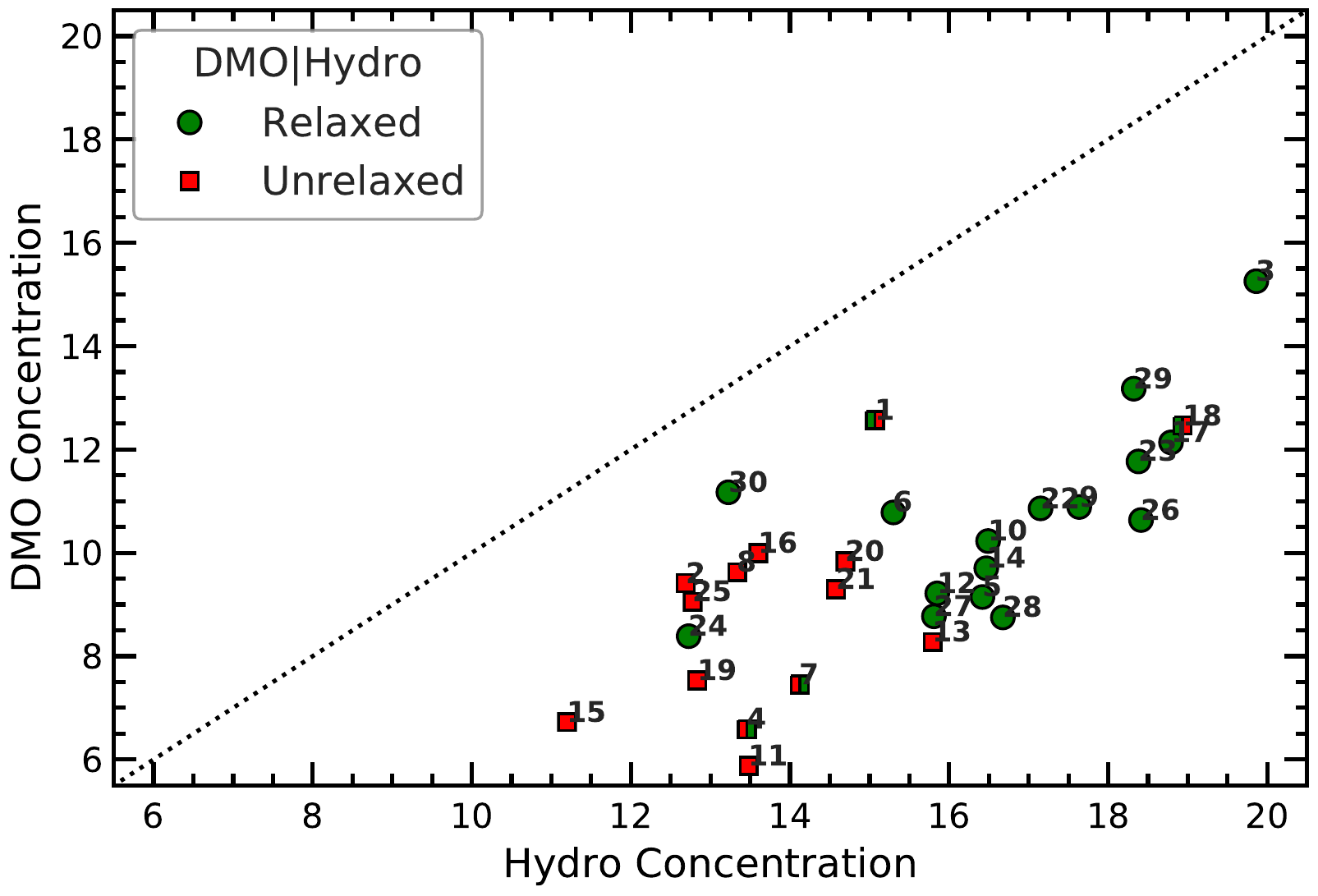}
	\vskip -.2cm
        \caption{ The concentrations,
          $c_{200}=R_{200}/r_{\textnormal{s}}$, of the 30 \auriga{}
          halos in the DMO and Hydro simulations, obtained by fitting
          an NFW profile to the DM distribution in each case. The
          points are green circles or red squares if the halos are relaxed or
          unrelaxed. The contraction of the DM halos in the Hydro
          simulations increases their concentration relative to the
          DMO case.}
    \label{fig:AurigaConc}
\end{figure}

We fit NFW profiles to the spherically averaged DM density profile of
our halos using least squares fitting in $\log r$ within the range
$R_{200}/100<r<R_{200}$. We find that the NFW profile provides a good
fit to the DMO halos, especially the relaxed ones; however it provides
a poorer description of the DM distribution in the Hydro simulations
\citep[see also e.g.][]{Schaller2016,CautunContract}. Nonetheless, for
completeness we calculate the best fitting NFW profile for the dark
matter halos in the Hydro simulation as well. In this case, because of
the poor fits, the inferred scale radius and concentration can
strongly depend on the radial range used for the fitting. The resulting
concentrations, $c_{200}=R_{200}/r_{\textnormal{s}}$, of the DMO and
Hydro halos are shown in Fig \ref{fig:AurigaConc}. The concentration
of the Hydro halos is systematically higher, indicating an increase in
DM density in the inner regions. It can also be seen that unrelaxed
halos typically have slightly lower concentration, in agreement with
previous studies \citep{Neto2007}.

\begin{figure}
	\includegraphics[width=\columnwidth]{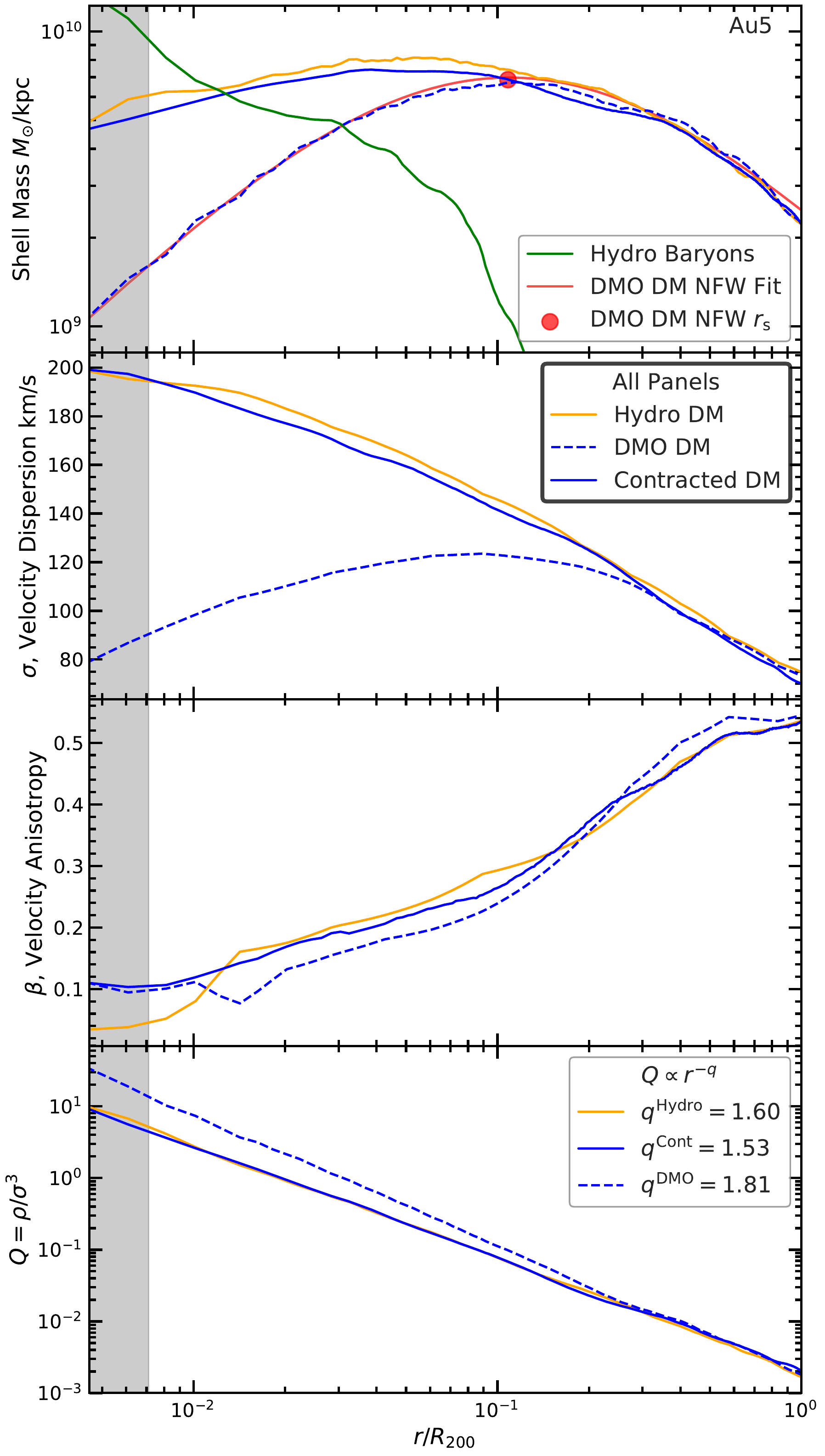}
	\vskip -.2cm
        \caption{ An illustration of the density, velocity dispersion
          and velocity anisotropy profiles of a DM halo (Auriga halo
          5) shown for the DMO (dashed blue) and the Hydro (solid
          orange) versions of the
          simulation. 
          Compared to the DMO case, the Hydro halo has a higher density in
          the central regions (top panel), along with an increased
          velocity dispersion (second panel). The third panel shows
          only small differences in the velocity anisotropy,
          $\beta$. The bottom panel shows the pseudo-phase-space
          density, $Q\left(r\right)=\rho/\sigma^{3}$, which we find is
          well fitted by a simple power law for both DMO and Hydro
          halos. Also plotted is the contracted DMO halo (solid blue),
          which was obtained by applying the method described in
          Sec. \ref{Subsec:SelfConsistentHalo}. This closely reproduces the Hydro
          halo. The grey shaded region corresponds to $r$ values
          below the convergence radius of the simulation
          \citep{Power2003}.  }
    \label{fig:HaloProp_r}
\end{figure}

The effects of contraction may be seen in more detail by comparing the
spherically averaged profiles of a halo in the DMO and Hydro
simulations. This is shown in Fig.~\ref{fig:HaloProp_r}, which
presents the shell mass, $M_{\mathrm{Shell}} = 4\pi r^{2}\rho$, the
velocity dispersion, $\sigma_{V}$, and the velocity anisotropy,
$\beta =1 - \sigma^2_T/\sigma^2_r$ (where $\sigma_t$ and $\sigma_r$
represent the tangential and radial velocity dispersions respectively)
for one of the relaxed halos, \Au5. The DMO density is scaled
by $ 1-f_{\textrm{Baryon}}$ to subtract the cosmic baryon fraction,
$f_{\textrm{Baryon}}=\Omega_{\textnormal{Baryon}} /
\Omega_{\textnormal{Matter}}$. As expected, the DMO halo density (top
panel) is well fitted by the NFW form, with the best-fit NFW profile
shown by the red solid curve. The velocity dispersion (top-middle
panel) of the DMO halo peaks just inside the scale radius, which
corresponds to the maximum of $M_{\mathrm{Shell}}$. The density at
each radius can be interpreted as a measure of the number of different
orbits at that radius, so the peak at the scale radius reflects the
relatively higher number of orbits that pass through this radius. The
velocity anisotropy, $\beta^{\textnormal{DMO}}\left(r\right)$, is nearly
isotropic in the centre and becomes more radially biased towards the
outskirts, again in agreement with previous studies
\citep{Tissera10DMRes_2010MNRAS.406..922T,Navarro10Diverse_2010MNRAS.402...21N}.
While all of our relaxed DMO halos conform to the NFW form, we see
significant scatter in their concentrations and variations in their
velocity dispersion and velocity anisotropy.

For the Hydro halo, we find a DM profile that is more centrally
concentrated (orange line in the top panel of
Fig.~\ref{fig:HaloProp_r}). This is due to response of the halo to the
baryonic distribution (green line), which is much more centrally
concentrated than in the DMO simulation  (in which, by construction,
the ``baryons'' 
have the same profile as the DM, but with a different
normalisation). The baryons deepen the central potential, compressing
the orbits of the DM particles inwards and significantly increasing the
DM density and total velocity dispersion in the central regions.
The velocity anisotropy, $\beta$, profile varies only slightly between
the DMO and Hydro halos, with the DMO halos typically having a
slightly more radially-biased velocity anisotropy between the scale
radius and $R_{200}$ (this is not the case for the Au-5 halo shown in
Fig. \ref{fig:HaloProp_r}), but there is significant halo-to-halo
scatter.

The bottom panel of Fig.~\ref{fig:HaloProp_r} shows the so-called
peudo-phase-space density,
$Q\left(r\right)=\rho/\sigma^{3}$. Surprisingly, in DMO halos this
quantity has been shown to closely follow a simple power law,
$Q\propto r^{-q}$, with a theoretically predicted slope, $q\sim 1.875$
\citep{BertschingerQ1985ApJS...58...39B}, that is consistent with our
results, $q\sim 1.84_{-0.07}^{+0.04}$. The origin of this relation
remains unclear, and whether it is a fundamental feature or a
dynamical `fluke' is debated in the literature
\citep[e.g.][]{LudlowQ2010MNRAS.406..137L,Navarro10Diverse_2010MNRAS.402...21N,LudlowQMill2011MNRAS.415.3895L,Arora19_QFluke_2019arXiv190403772A}. We
find that the Hydro halos also conform to this power law (in agreement
with \citealt{Tissera10DMRes_2010MNRAS.406..922T}), with similar
scatter but with a shallower slope
$Q_{\textnormal{Baryon}}\sim 1.62_{-0.08}^{+0.08}$. We leave this
interesting observation for future work.

\subsection{Orbital Phase Space}
\label{Subsec:OrbitalPhaseSpace} 

As we discussed in the introduction, we are interested in describing
DM halos in terms of their action distribution, \FJ{}. This provides a
complete description of the orbits of particles in the halo, which can
be used as the blueprint to reconstruct various halo properties, as
we shall see in the next section.

We model halos as spherically symmetric distributions for which
the gravitational potential, $\Phi\left(r\right)$, is related to the
total density profile, $\rho\left(r\right)$, by:
\begin{equation} \label{eq:Potential}
	\Phi\left(r\right)=-4\pi G\left(\frac{1}{r}\int_{0}^{r}r'^{2}\rho\left(r'\right)\mathrm{d}r'+\int_{r}^{\infty}r'\rho\left(r'\right)\mathrm{d}r'\right),
\end{equation}
where $G$ is Newton's gravitational constant.  Spherical symmetry
reduces the number of actions needed to describe each orbit to two as
the third action is identically zero and the orbit stays in a
plane between its pericentre, $r_{\textnormal{min}}$, and apocentre,
$r_{\textnormal{max}}$.

The two nonzero actions are the specific angular momentum, $L$, and
the radial action, $J_{r}$, given by:
\begin{equation} \label{eq:ActionEq}
\begin{aligned}
L &= \left|\boldsymbol{r}\times\boldsymbol{v}\right|=r v_{\textnormal{t}}\\
\Jr &=\frac{1}{\pi}\int_{r_{\textnormal{min}}}^{r_{\textnormal{max}}}  v_{r}\left(r\right)\textnormal{d}r
\end{aligned}
\end{equation}

An alternative IoM commonly used in dynamical modelling is the
(specific) energy, $E$, defined as:
\begin{equation}
    E = \frac{1}{2}\left|\textnormal{\textbf{v}}\right|^{2}+\Phi\left(r\right)\\
\end{equation}
While convenient to calculate, $E$ is not an adiabatic invariant. The
energy distribution function, \FE{}, is therefore expected to differ
systematically between the DMO and the Hydro simulations, whereas \FL{} and
\FJr{} are expected to remain approximately the same. Note that in
this paper all distributions, $F$, are normalised to integrate to 1.
 
\begin{figure}
	\includegraphics[width=0.9\columnwidth]{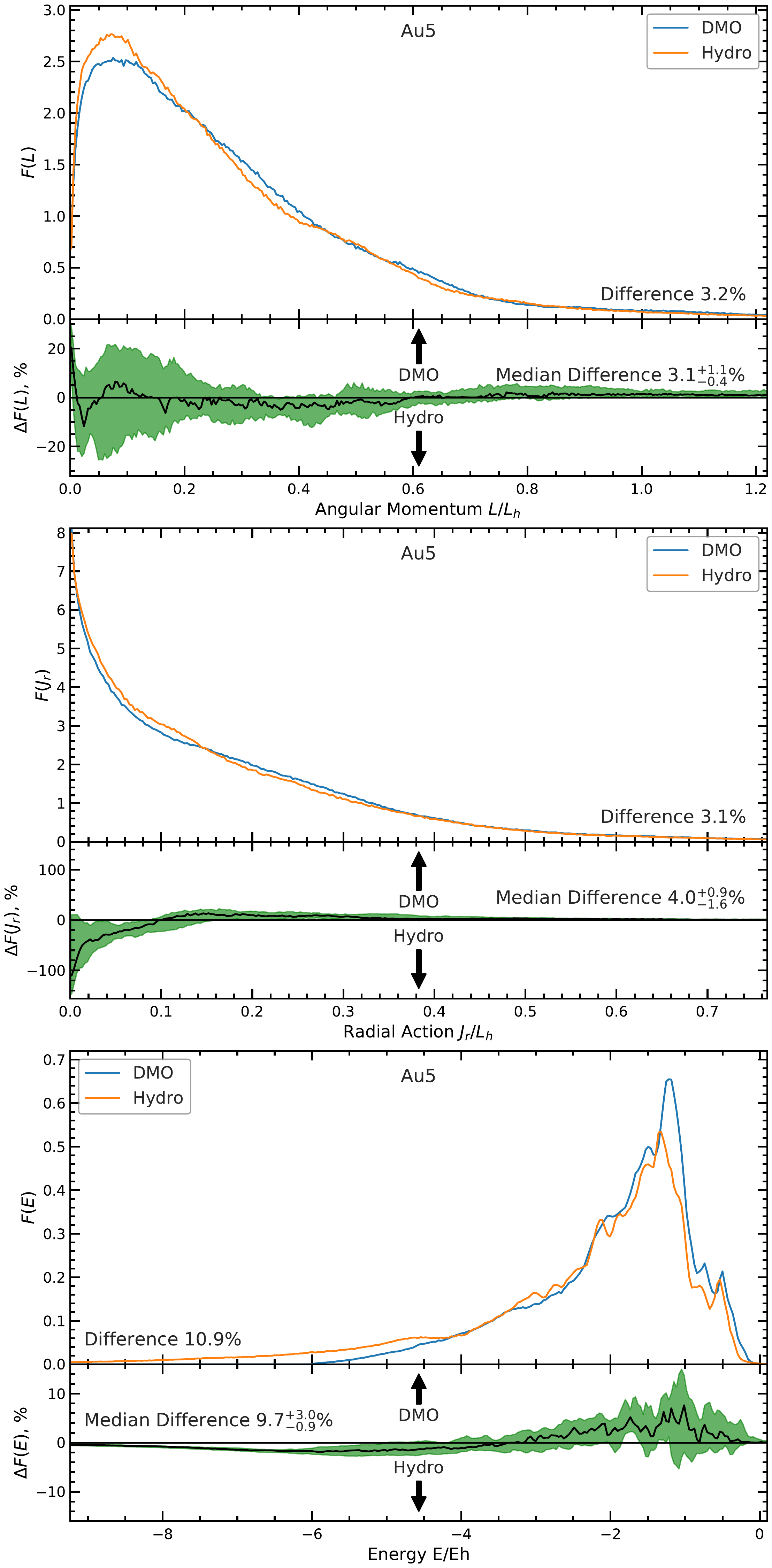}
	\vskip -.2cm
        \caption{ The distributions of angular momentum, $L$, radial
          action, \Jr{}, and energy, $E$, of the DM particles in the
          Hydro and DMO simulations for an example relaxed halo, Au5
          (top subpanels). In general, we find small differences
          between the distribution functions of the adiabatic
          invariant actions, \FL{} and \FJr{}, in the DMO and Hydro
          cases. The distribution of the non-adiabatic invariant
          energy, \FE{}, shows larger differences. To check if these
          differences are systematic, the bottom subpanels show the
          median (black solid line) and 68 percentiles (green shaded
          region) of the difference between the DMO and Hydro
          distributions, $\Delta F = F_{\text{DMO}}-F_{\text{Hydro}}$,
          for all relaxed \auriga{} halos. To compare different halos,
          the orbital values are scaled to be independent of halo mass
          (for further details see the main text). The DM energy
          distributions (bottom panel) are most affected by the
          presence of baryons, with about 10$\%$ of the particles
          changing energy. These are mainly inner DM particles
          shifting to lower energies in the deeper Hydro
          potential. The distributions of the actions, $L$ and
          $J_{\text{r}}$, experience smaller changes, $3\%$ and $4\%$
          respectively.  }
 \label{fig:OrbitalDist}
\end{figure}

Here the distributions are found for each halo by selecting, from the
centre outwards, the same number of DM particles for each DMO and Hydro counterpart halo, contained 
 within $R_{200}$ of the Hydro halo. In
general, halos in the DMO and Hydro simulations are well matched.
However, the stochastic nature of galaxy formation, as well as the small inherent numerical efffects, cause small differences in the distributions of DM
particles. On average, we find that $\sim90\%$ of the DM particles
within $R_{200}$ in the Hydro case are also found within $R_{200}$ in
the DMO case.  We have checked that differences in the halos' orbital
distributions discussed in this study are not caused by unmatched DM
particles between the Hydro and DMO cases; distributions of matched
particles differ by similar amounts.

To compare the distributions of different mass halos, the IoMs (of both
the Hydro and DMO halos) are rescaled to give values that are independent
of the host halo mass (see
\citealt{Zhu2016,CallinghamMass2019MNRAS.484.5453C}). The actions, $L$
and \Jr, are normalised by the characteristic angular momentum of a
circular orbit at $R_{200}$, $L_{\text{h}}=\sqrt{GM_{200}
  R_{200}}$. The energy is similarly normalised by this orbit's energy,
$E_{\text{h}}=GM_{200}/R_{200}$.

In Fig.~\ref{fig:OrbitalDist} the distributions of $L$, \Jr{} and $E$
for one example halo (Au-5) are shown in the top subpanels. The lower
subpanels show the difference between the distributions in the DMO and
Hydro cases, $\Delta F = F_{\text{DMO}}-F_{\text{Hydro}}$, for all of
the relaxed level 4 \auriga{} halos; the solid line is the median and
the shaded region indicates the 68 percentiles of the distribution.
To estimate the difference between the various distributions, we
calculate the overall difference, $\Delta$, which is effectively the
fraction of DM particles whose IoM are distributed differently between
the Hydro and DMO cases. This is defined as:
\begin{align}
    \Delta_{\rm{X}} & = 
    \frac{1}{2}\int \left| F_{\text{DMO}}(X)-F_{\text{Hydro}}(X) \right| \ \textnormal{d}X 
    \nonumber \\
        & \equiv \frac{1}{2}\int\left| \Delta F \left( X \right)
          \right| \ \textnormal{d}X, 
    \label{eq:diff_1D_distributions}
\end{align}
where $X$ denotes the IoM under consideration, either $L$,
$J_r$ or $E$. With this normalisation, $\Delta_{\rm{X}}=1$ when the
distributions are completely different. 

The distributions \FL{} (top panel) and \FJr{} (middle panel) are
similar to those found in previous simulations
\citep{PontzenEntropy2013MNRAS.430..121P}. Between the DMO and Hydro
simulations there is a small, seemingly stochastic difference, in
angular momentum ($\Delta_{L}\sim3\%$) at low $L$. The difference in
\Jr{} is also small, $\Delta_{J_{r}}\sim4\%$, but systematic, with a
slight increase towards low $J_{\text{r}}$ for the DM particles in the
Hydro case. The energy distributions, $F\left(E\right)$ (bottom
panel), have distinct peaks and features unique to the individual halo
that are not present in the other IoM. These are remnants of a complex
merger history, with similar features in the counterpart halo.  The
energy distributions are most affected by contraction with
$\Delta_{E}\sim 10\%$ as the deeper central potential of the Hydro
halo reduces the energy of the inner DM particles.

We saw that the \Jr{} and $L$ one-dimensional distributions are
roughly conserved between the Hydro and DMO simulations. But what
about the joint two-dimensional \FJrL{} distribution? Is it also
conserved? This question is relevant since we find correlations
between \Jr{} and $L$, as illustrated in the top panel of
Fig.~\ref{fig:Dist2D}. These correlations vary between halos and
potentially encode important information about the halo's density and
velocity profiles. To find the answer, we calculate the differences in
the \FJrL{} distributions between the DMO and Hydro cases; similarly
to Eq.~\eqref{eq:diff_1D_distributions}, the action difference, \Da{},
is defined as:
\begin{equation}
    \Da=\frac{1}{2}\intop\left|F_{\textnormal{DMO}}\left(J_{r},L\right)-F_{\textnormal{Hydro}}\left(J_{r},L\right)\right| \; \textrm{d}L\textrm{d}J_{r} \; .
\end{equation}
For relaxed halos, $\Da \sim 8 \pm1\%$. This is larger than the
differences in the one-dimensional distributions, but nonetheless it
is still rather small indicating that the joint distribution is
roughly invariant too. The value of $\Da$ is used in the appendix~\ref{Appendix:Adibiatic} to study the extent to which differences in action distribution are related to differences between the
contracted DMO halos and their Hydro counterparts.


\begin{figure}
	\includegraphics[width=\columnwidth]{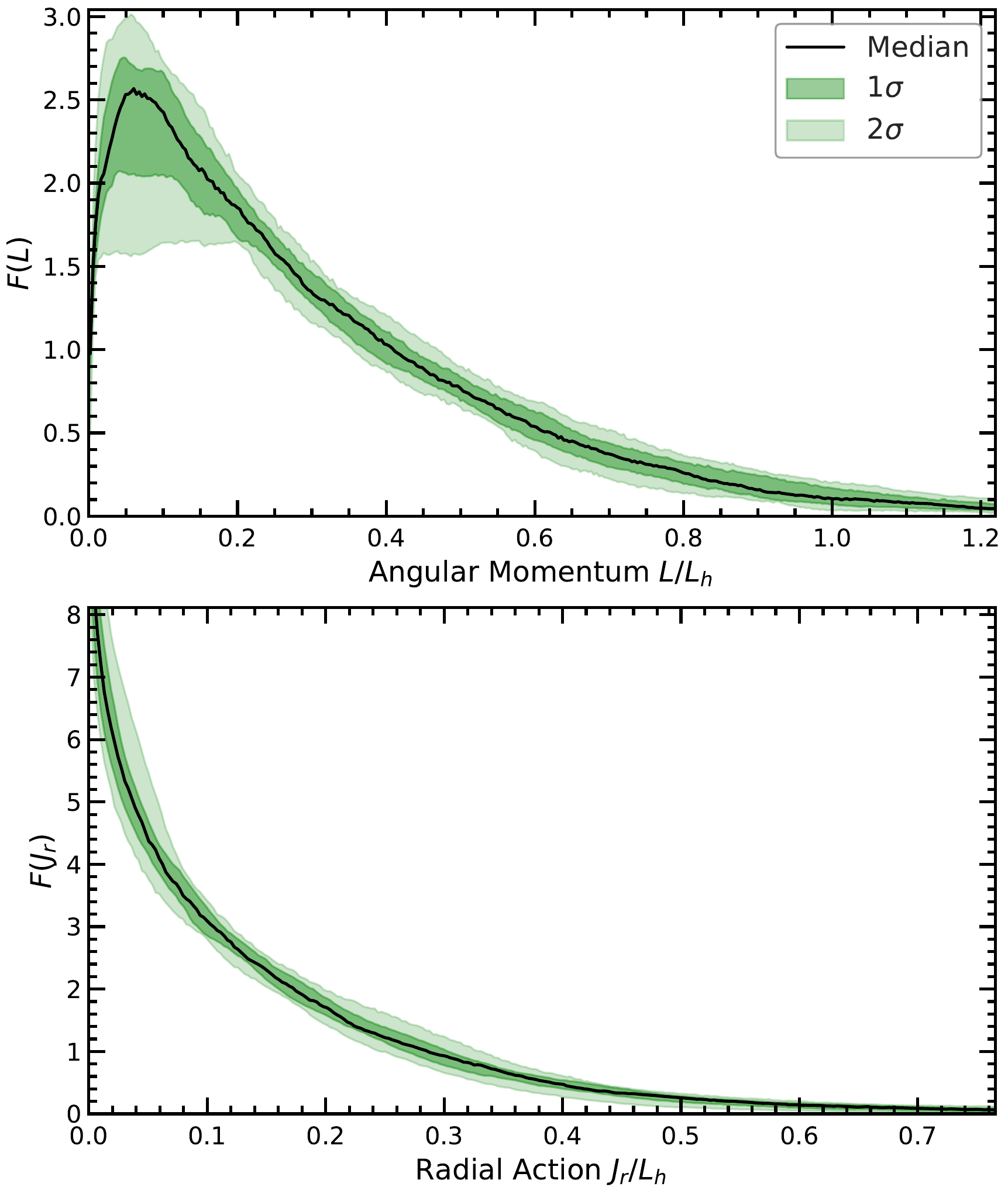}
	\vskip -.2cm
        \caption{The distributions of angular momentum, $L$, and
          radial action, \Jr{}, of DM particles in the DMO simulation
          for our sample of relaxed halos.  The black solid line shows
          the median of our sample and the green shaded region the 68
          percentile and full halo-to-halo scatter.  To compare halos,
          $L$ and \Jr{} are scaled to be independent of halo mass (for
          details see the main text).  }
 \label{fig:OrbitalDist_halo_scatter}
\end{figure}

Fig.~\ref{fig:OrbitalDist_halo_scatter} shows the action distributions
\FL{} and \FJr{} of our relaxed \auriga{} halo sample. While the
individual action distributions have qualitatively similar form,
differences in the peak of the distributions suggest
object-to-object scatter in the DFs, which could arise from different
halo formation histories. This is to be expected as NFW profiles fit
the majority of halos very well, but the concentration and
$\beta\left(r\right)$ profiles vary from halo to halo. We leave the
precise characterisation of these distributions and a potential
concentration parameterisation to future work. Here we investigate
the effects of halo-to-halo variation by calculating the contracted DM
halo using multiple \FJ{} distributions.

\section{Constructing the Halo from Particle Orbits}
\label{Sec:HaloConstruction} 

In the previous section we calculated the distribution of DM particle
orbits as described by their spherical actions distribution,
\FJrL{}. We now calculate the individual orbits in physical space to
find their contribution to the structure of the DM halo.  We will use
this information in the next subsection where we construct the
physical properties of the DM halo, such as its density and velocity
dispersion profiles, by summing over the orbital distribution, \FJrL.
Instead of considering a particle as a point contribution to the halo,
we consider the physical contribution of its orbit sampled uniformly
in phase, i.e. we consider the contribution of the particle spread
around its orbit in time. The radial distribution of an orbit,
$\FrJrL$, is defined as the proportion of time that orbit spends at
radius $r$, normalised so that it integrates to unity.  This is
approximately:
\begin{equation}
    \FrJrL\approx \frac{2}{T|_{\Jr,L}}\frac{1}{v_{r}\left(r\right)|_{\Jr,L}}
    \;,
\end{equation}
where $T$ is the radial time period and $v_{r}$ is the radial
velocity \citep[see][]{HanOPDF_2016MNRAS.456.1003H}.
However, this is only an approximation and great care is
needed at the endpoints where $v_{r}\xrightarrow{}0$. For a more
detailed derivation and further details please see
Appendix~\ref{Appendix:Orbits}.  The density can then be reconstructed
by integrating over the distribution of these orbits:
\begin{equation}
    \label{Eq:DensityOrbits}
    \rho\left(r\right)=\frac{M_{\textnormal{DM}}}{4\pi r^{2}}\iint
    \FrJrL \FJrL \ \textrm{d} \Jr \textrm{d} L, 
\end{equation}
where $M_{\textnormal{DM}}$ is the total mass of the DM halo. 

When contracting a DMO halo to account for its baryon distribution,
the cosmic baryon fraction must be removed in order to obtain the
correct DM halo mass. That is, the mass of the DM halo in the DMO case
is given by $\left(1-f_{\textrm{Baryon}}\right)$ times the total halo
mass.  When constructing the halo, \FJrL{} must include all DM
particles within (and orbits calculated up to) $3R_{200}$ to ensure
all significant contributions to the halo are included.

In practice, it is simpler first to construct orbits from a given
$\left(E,L\right)$ pair and a potential, $\Phi \left(r \right)$. The
\FEL{} distribution is derived from \FJrL{}, given a potential
$\Phi$. This can be evaluated numerically using the \Jr{} calculated
from each $\left(E,L\right)$ pair as:
\begin{equation}
    \FEL=\FJrL\frac{\mathrm{d}J_{r}}{\mathrm{d}E}
    \;
\end{equation}{}
where \FJrL{} is evaluated by interpolating the halo action distribution. We can now rewrite Eqn.~\eqref{Eq:DensityOrbits} in terms of the energy
and angular momentum distribution to obtain
\begin{equation}
    \label{Eq:DensityOrbitsEL}
    \rho\left(r\right)=\frac{M_{\textnormal{DM}}}{4\pi r^{2}}\iint \FrEL \FEL \ \textrm{d} E \textrm{d} L
    \; .
\end{equation}

\begin{figure}
    \centering
	\includegraphics[width=\columnwidth]{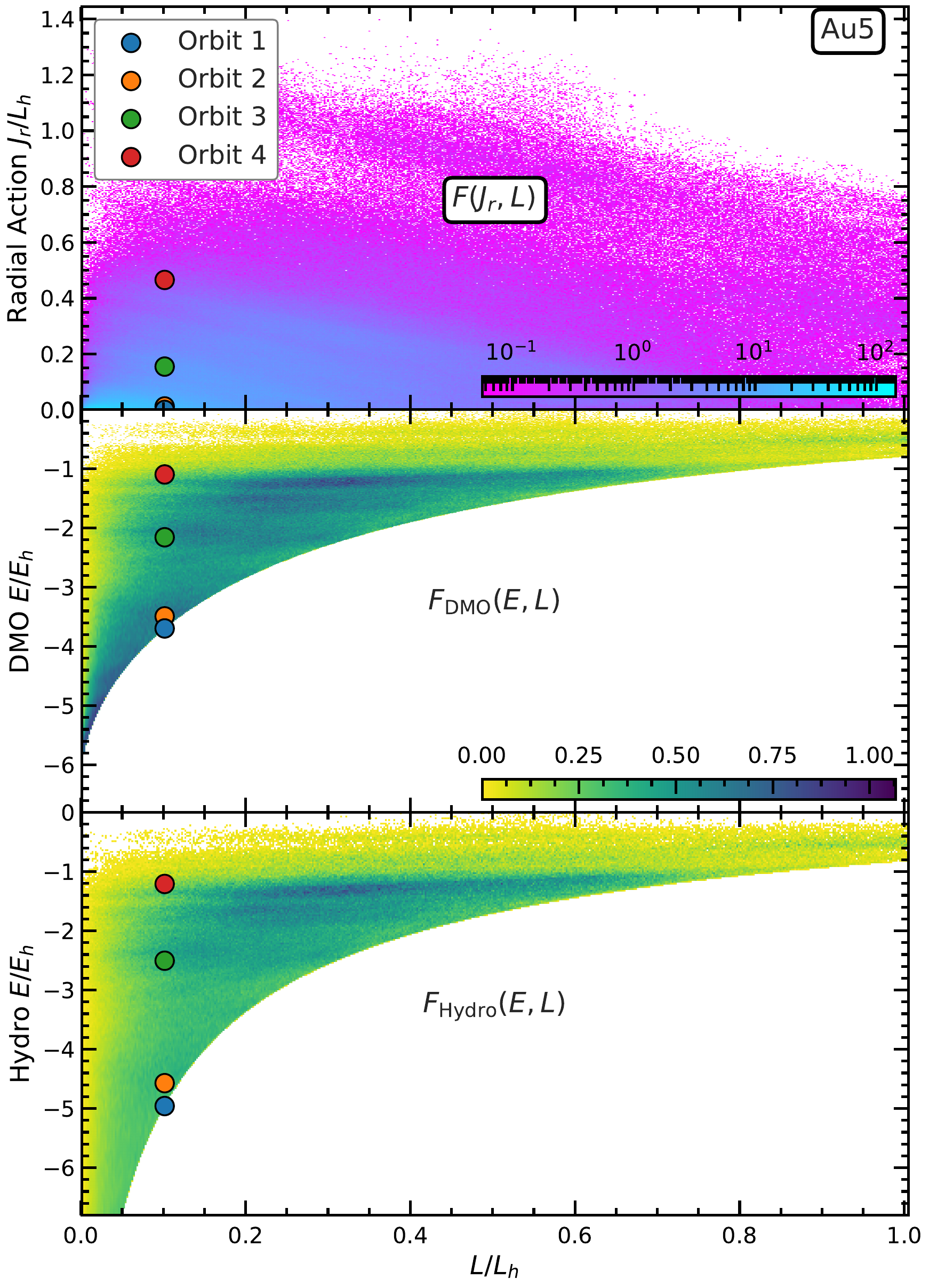}
	\vskip -.15cm
        \caption{The 2D distribution, \FJrL{}, of radial action,
          \Jr{}, and angular momentum, $L$, of the DM particles in the
          DMO simulation of the \Au5 halo (top panel).  Given a gravitational
          potential, \FJrL{} can be used to calculate the 2D
          distribution, $F\left(E,L\right)$, of energy, $E$, and
          $L$. The result is illustrated in the centre and bottom
          panels, which show $F\left(E,L\right)$ for the DMO and Hydro
          simulations respectively. The deeper potential in the Hydro case
          leads to overall lower energy orbits. To better illustrate
          the transformation, the coloured symbols show four orbits
          selected to have the same $L$, but different \Jr{} values.
          The radial profiles of these orbits are shown in
          Fig.~\ref{fig:OrbitRadDist}.  The actions are given in units of $L_{\text{h}}=\sqrt{GM_{200}}$ and Energy in $E_{\text{h}}=GM_{200}/R_{200}$.
          }
    \label{fig:Dist2D}
\end{figure}

We calculate the orbits using a $500^{2}$ grid in $(E,L)$ space. We find
that this grid size is a good compromise between computational time
and the sufficiently high orbit density needed to recover a smooth
halo profile. We have experimented with different methods for defining
the $(E,L)$ grid and have selected the one that gives accurate results
for the smallest grid size. This is obtained by first choosing 500 $L$
values, evenly spaced in the cumulative \FL{} distribution. Then, for
each $L$ bin, we select 500 $E$ values evenly spaced on the allowed
phase space, that is in the interval
$\left[E_{\textnormal{Circ}}\left(L\right),0\right]$. By doing so, we
neglect unbound particles, i.e. particles with positive total energy,
$E>0$. However, there is only a small fraction of such particles
($\sim 0.05\%$; see Fig.~\ref{fig:Dist2D}) and, in practice,
excluding them makes no difference.

We illustrate the transformation from $\left(\Jr,L\right)$ space to
$\left(E,L\right)$ space in Fig.~\ref{fig:Dist2D}. The top panel shows
the distribution, \FJrL{}, of the \Au5 halo in the DMO simulation.
The bottom two panels show the distribution, \FEL{}, for the DMO and
Hydro simulations respectively, which have been calculated from the action DF shown in the top panel
using the actual gravitational potential measured in each of the two
cases. The \FEL{} distributions are bounded on the lower right edge by circular orbits,
which have the minimum energy possible for a given angular
momentum. Compared to the DMO case, the Hydro simulation is
characterised by more lower energy orbits, a manifestation of the
deeper potential well of the Hydro halo.

\begin{figure}
    \centering
	\includegraphics[width=\columnwidth]{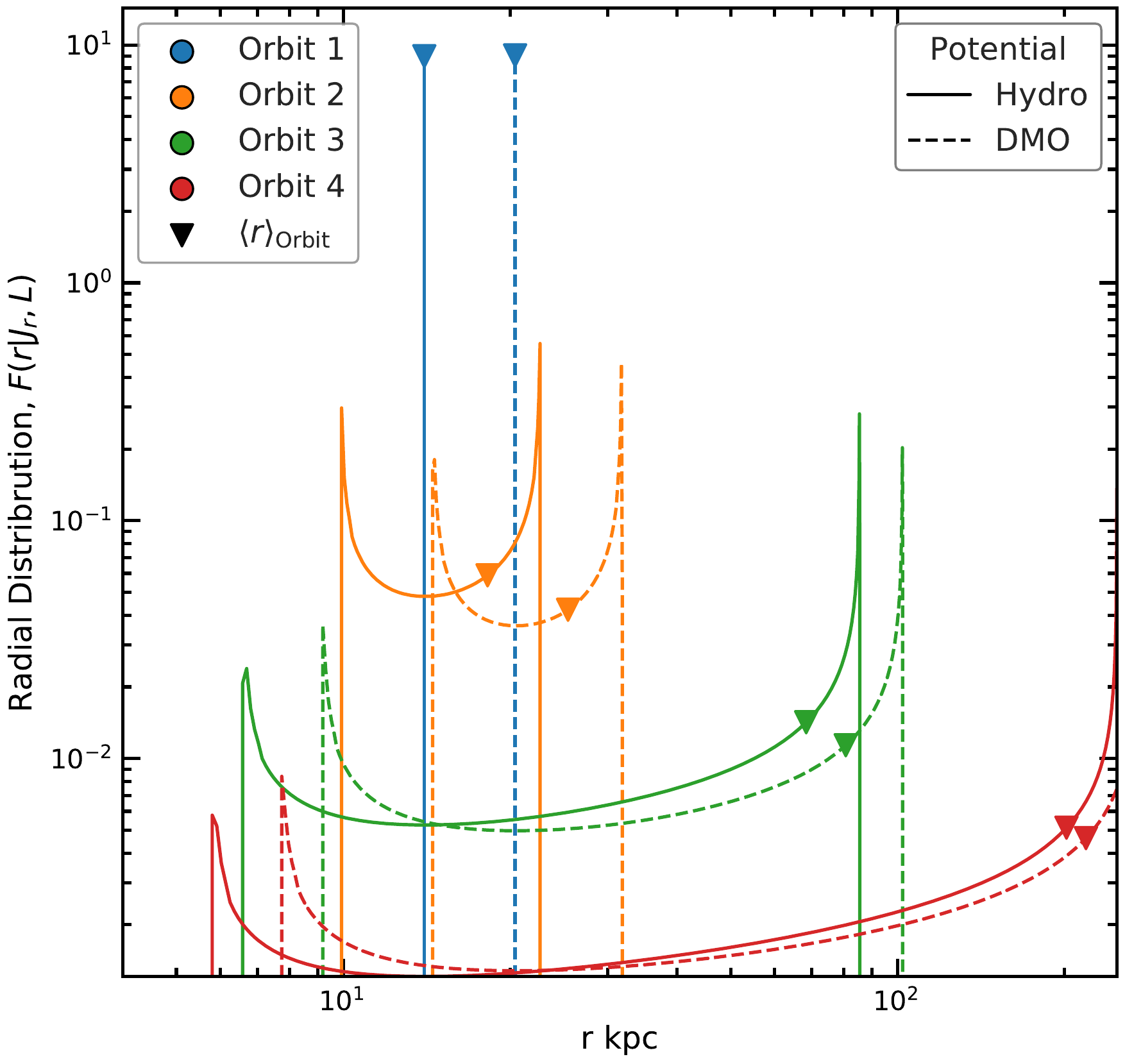}
	\vskip -.15cm
        \caption{
           The radial distribution, \FrJrL,  for four different
          orbits (each shown by a different colour). This is equivalently the 
          fraction of time that a particle on
          orbit $(\Jr,L)$ spends at a given radius, per kpc.  The orbits have
          the same angular momentum, $L$, but increasing radial
          action, \Jr{} (from Orbit 1 to Orbit 4, where Orbit 1 is
          circular -- see coloured symbols in
          Fig.~\ref{fig:Dist2D}). We show the orbits for the
          gravitational potential of the \Au5 halo in the DMO (dashed
          lines) and Hydro (solid lines) cases. The triangles show the
          median radius of each orbit. The deeper Hydro potential
          pulls the orbits to lower radius, affecting the more
          circular orbits the most.  }
    \label{fig:OrbitRadDist}
\end{figure}

To gain a better understanding of how a given orbit, $(\Jr,L)$, changes
between the DMO and Hydro potentials, we select 4 orbits with the same
angular momentum, $L=0.12L_{h}$, and increasing radial action,
$ \Jr=[0,400, 5000,15000] \, \textnormal{km}\, \textnormal{kpc}\,
\textnormal{s}^{-1}$.  These orbits are shown as coloured symbols in
Fig.~\ref{fig:Dist2D}. The lower the \Jr{} of the orbit, the larger
the decrease in energy from the DMO to the Hydro potential, as can
be determined from the bottom two panels of Fig.~\ref{fig:Dist2D}. 

The change in energy of the orbits between the DMO and Hydro
potentials is accompanied by a pronounced change in the radial range
associated with a $(\Jr,L)$ orbit.  This is illustrated in
Fig.~\ref{fig:OrbitRadDist}, which shows the fraction of time,
\FrJrL{}, that a particle on orbit $(\Jr,L)$ spends at different
distances from the halo centre. The figure shows the same four orbits
highlighted in Fig.~\ref{fig:Dist2D}. To help interpret the plot, each
orbit in Fig.~\ref{fig:OrbitRadDist} is marked with a triangle symbol,
which shows the median radial position of the orbit: a particle spends
half its orbital time at farther distances than this. Orbit~1 is
circular and lies at the scale radius of the DMO halo. With increasing
\Jr{} the orbits gain radial kinetic energy and become more radial, so
their median radial position occurs further out from the circular
radius. The orbits spend most of their time at the endpoints,
i.e. pericentre and especially apocentre (note the logarithmic
y-axis), while they spend the least amount of time at the circular
radius for their given angular momentum where $v_{r}$ is maximal.

Adding baryons deepens the potential well and the orbits are pulled
inward, leading to a compression of the DM halo. This can be seen by
comparing the DMO orbits (dashed lines) with the Hydro ones (solid
lines). The more circular orbits are compressed the most, with
fractional decreases in the median radius of orbits from 0.7 for
Orbit~1 to 0.9 for the most radial Orbit~4. This is agreement with the
suggestion that radial orbits `resist' compression
\citep{SellwoodContract_2005ApJ...634...70S,Gnedin2004}.

\subsection{ Finding a Self-consistent Halo }
\label{Subsec:SelfConsistentHalo} 

\begin{figure}
	\includegraphics[width=\columnwidth]{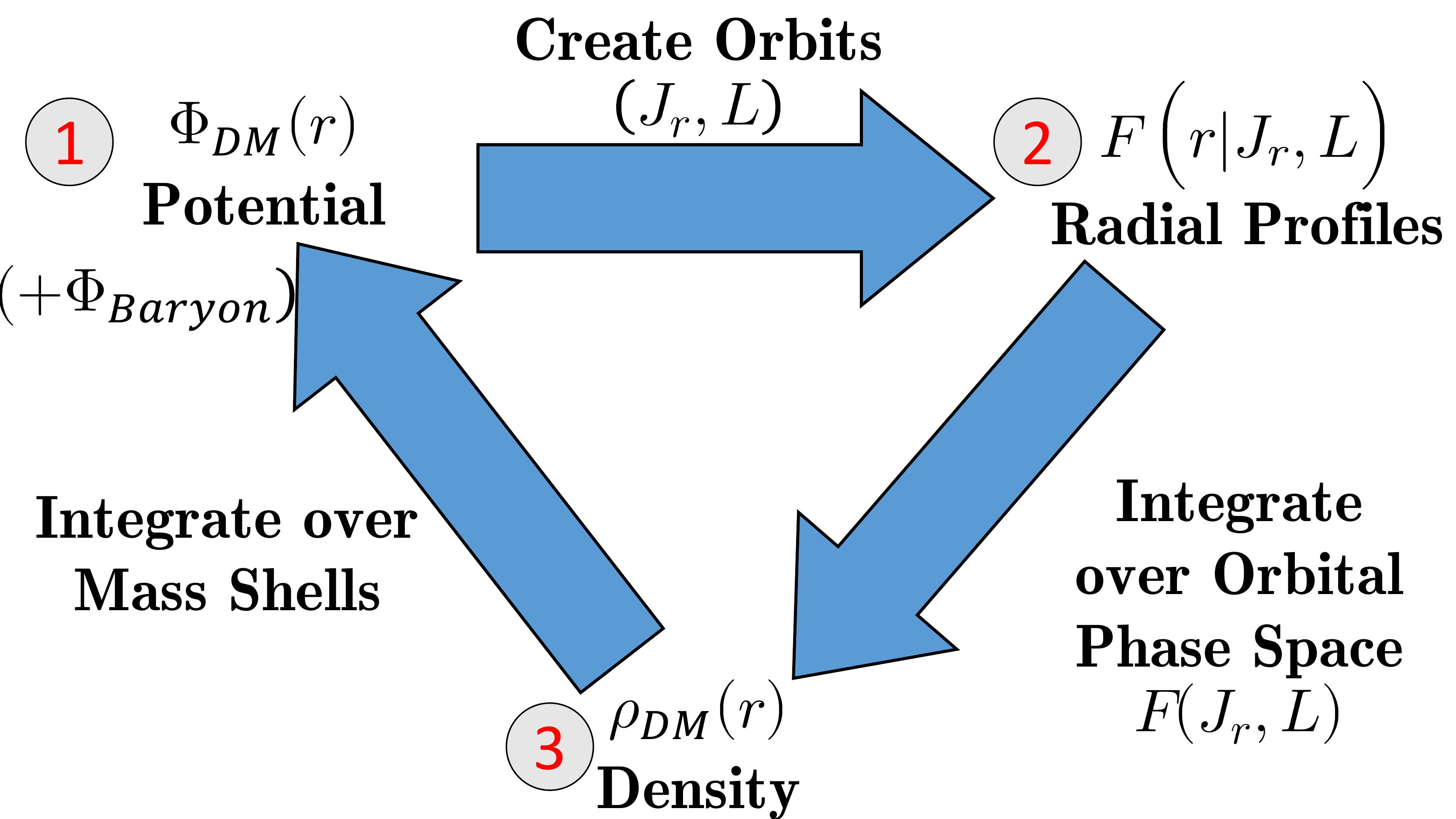}
	\caption{ Flowchart of an iterative scheme to calculate a halo
          density profile starting from its action distribution,
          \FJrL. The method proceeds as follows: 1)~using a trial 
          gravitational potential for the DM, $\Phi_{\rm DM}$,
          calculate the radial range, \FrJrL{}, of each orbit,
          $\left(\Jr,L\right)$; 2)~integrate over all orbits to
          calculate the DM density profile, $\rho_{\mathrm{DM}}$; 
          3)~use the inferred DM density to update the DM potential,
          $\Phi_{\rm DM}$; and repeat from step~1) until convergence
          is achieved. If required, an additional baryon potential
          $\Phi_{\mathrm{Baryon}}$ can be added in step~1) to find a
          contracted halo.  }
	\label{fig:Flow}
\end{figure}

Our aim is to construct a DM halo in physical space, inferring the
density and velocity profiles solely from the DM action distribution, 
\FJrL.  In the previous Section we showed that given a fixed
potential, $\Phi$, we can obtain the DM density profile,
$\rho_{\textnormal{DM}}\left(r\right)$, from the action DF by
calculating the radial distribution, \FrJrL{}, of individual orbits
that is then integrated over \FJrL{} to obtain the overall radial
distribution of DM particles (see Eqn. \ref{Eq:DensityOrbits}).

To obtain the true halo density profile we need to know the total
gravitational potential, $\Phi$, of the baryonic and DM
components. The challenge arises from the fact that the DM
gravitational potential needs also to be calculated from the action
distribution. Here we describe how this can be done in a
self-consistent way using an iterative approach. We first make an
initial guess for the potential which, at each iteration, is updated
to a value that is ever closer to the true potential.

Our approach is illustrated in Fig.~\ref{fig:Flow} and proceeds as
follows. First a sensible trial potential,
$\Phi_{\textnormal{DM}}^{0}$, is chosen, for example, the potential of
an NFW halo of average concentration for the target halo mass. When
considering the Hydro halo, we typically choose the DM potential from
the counterpart DMO halo since this achieves faster convergence. We
sum the DM and baryon\footnote{The baryon potential is kept fixed and
  is an input to the method, e.g. the potential from the stellar
  distribution of an \auriga{} halo or of the MW. The method applies
  to DMO simulations too, in which case the baryon potential is
  obtained as the cosmic baryon fraction multiplied by the total
  potential measured in the simulation. The same result is obtained if
  instead we take a null baryon potential and assume that the DM
  constitutes 100\% of the mass in the DMO simulation.}  potentials to
obtain the total potential. The DM density is then calculated using
Eqn.~\eqref{Eq:DensityOrbits}, which, in turn, is used to determine
the updated DM potential. This is used as the input potential for the
next iteration step, which is repeated until convergence is
achieved. The convergence criterion is satisfied when the change in DM
density between two iterations is small enough. This is quantified in
terms of
\begin{equation}
\begin{aligned}
    \Dr^{\rm total} & = \left(\log\left(\frac{100}{3}\right)\right)^{-1}\intop_{R_{200}/100}^{R_{200}/3}\left|\Delta_{\rho}\left(r\right)\right| \ \textrm{d}\log r
    \; ,
\end{aligned}
\end{equation}
where $\Delta_{\rho}\left(r\right)$ is defined as the fractional
difference between two density profiles,
\begin{equation}
    \Delta_{\rho}\left(r\right)=2 \ \frac{ \rho_{2}(r)-\rho_{1}(r)}{\rho_{2}(r)+\rho_{1}(r)}
    \; .
\end{equation}
The quantity $\Dr^{\rm total}$ characterises the integrated difference
between two density profiles in the inner region of the halo, that is
for $r\in[\tfrac{1}{100},\tfrac{1}{3}]R_{200}$.  When running the
iterative approach without a convergence criterion, we find that
$\Dr^{\rm total}$ reaches a constant small value,
$\Dr^{\rm total} \in\left[0.01,0.005\right]\%$ (the exact value varies
from halo to halo). The final equilibrium state seems to be reached
inside out, with the outskirts of the halo converging somewhat more
slowly than the inner parts. Based on this, we choose to stop the
iterative procedure to determine the potential when
$\Dr^{\rm total}<0.02\%$.

We have tested the method by applying it to relaxed \auriga{} halos in
both the DMO and Hydro simulations. For example, we measured the
\FJrL{} distribution for a DMO halo, which was then used to recover
that halo's density profile starting from an initial potential given
by an NFW halo of average concentration for its mass.  When compared
with the `true' DM halo profile from the simulation we find very good
agreement: the density is typically recovered to within${\sim}2\%$ within $R_{200}/2$ with increasing scatter of $5\% to 10\%$ towards the outskirts of the halo. Differences mainly arise from assuming steady
state halos in which particles are uniformly spread in phase along
their orbits. However, recently accreted material and substructures do
not satisfy this assumption and can lead to differences between the
density profile measured in the simulations and that predicted by our
method.

\subsubsection{Scaling the action distribution to halos of different masses}
\label{SubSec:Scaling}

In this section we show how to scale our results from \auriga{} halos
to halos of arbitrary mass. We do this within the context of our
method for generating a halo from a given \FJrL{} distribution. The
goal is to take the \FJrL{} distribution measured for a halo of total
mass, $M_{200}^{\mathrm{initial}}$, and rescale it so that it can be
used to predict the profile of a target halo with total mass,
$M_{200}^{\mathrm{target}}$. For this, we exploit the observation that
DM halos, at least in DMO simulations, are universal when scaled
appropriately \citep[for more details see the discussion
in][]{Li2017,CallinghamMass2019MNRAS.484.5453C}. As we saw in
Fig.~\ref{fig:OrbitalDist}, the action distribution for the DMO and
Hydro simulations are very similar so we expect the universality to
apply to the action distribution not only in the DMO case, but also
when including a baryonic component. 

As we are interested in matching the total mass of a target halo with
a fixed given baryonic profile, we are only free to rescale the mass
of the DM halo, not that of the baryonic component. We define the mass
scaling factor,
$\lambda=M_{\rm 200;\ DM}^{\mathrm{target}}/M_{\rm 200;\
  DM}^{\mathrm{initial}}$, which is the ratio between the DM mass
enclosed within $R_{200}$ for the target and initial halos,
respectively. For DMO halos, we can rescale the initial halo to the
target one by rescaling the positions and velocities by
$\lambda^{1/3}$, and the energy and actions by $\lambda^{2/3}$. For
Hydro halos, rescaling the position, velocities and energy using the
same procedure is not a good strategy, especially in the inner halo
regions, where the universality of halos is degraded by the presence
of baryons. However, as we discussed earlier, this is not the case for
the actions, which scale as in the DMO case.

The rescaled action is given by
\begin{equation}\label{Eq:ScaleDens}
  \begin{aligned}
F'\left(\Jr,L\right)\equiv     F^{\mathrm{target}}\left(\Jr,L\right)=&\lambda^{-4/3} F^{\mathrm{initial}} \left(\Jr/\lambda^{2/3},L/\lambda^{2/3}\right),
  \end{aligned}
\end{equation}
where $F^{\mathrm{target}}$ and $F^{\mathrm{initial}}$ denote the
action distribution in the target and original halos respectively,
and the $\lambda^{-4/3}$ multiplication factor ensures that
the new distribution integrates to unity. We then use these new
actions, $F'\left(\Jr,L\right)$, as input to the method for
constructing the halo density profile described in
Section~\ref{Subsec:SelfConsistentHalo}.

The total mass, $M_{200}^{\mathrm{new}}$, of the resulting rescaled
halo is close to the target mass, $M_{200}^{\mathrm{target}}$, but
there can be small differences of order a few percent.  These are
present when baryons are included since the baryonic distribution can
either contract or expand the DM distribution and thus introduce small
variations in the total mass within $R_{200}$. We account for these
small differences by applying again the rescaling method, with the
actions now rescaled by a new factor,
$\lambda'=M_{\rm 200;\ DM}^{\mathrm{target}}/M_{\rm 200;\
  DM}^{\mathrm{new}}$, which is typically very close to one. Using the
new actions, we calculate again the halo density profile and its total
mass, $M_{200}^{\mathrm{new}}$, repeating the procedure until
convergence to the target halo mass is achieved.

\subsection{ Contracting Auriga Halos }
\label{Subsec:ContractAuriga} 

\begin{figure}
	\includegraphics[width=\columnwidth]{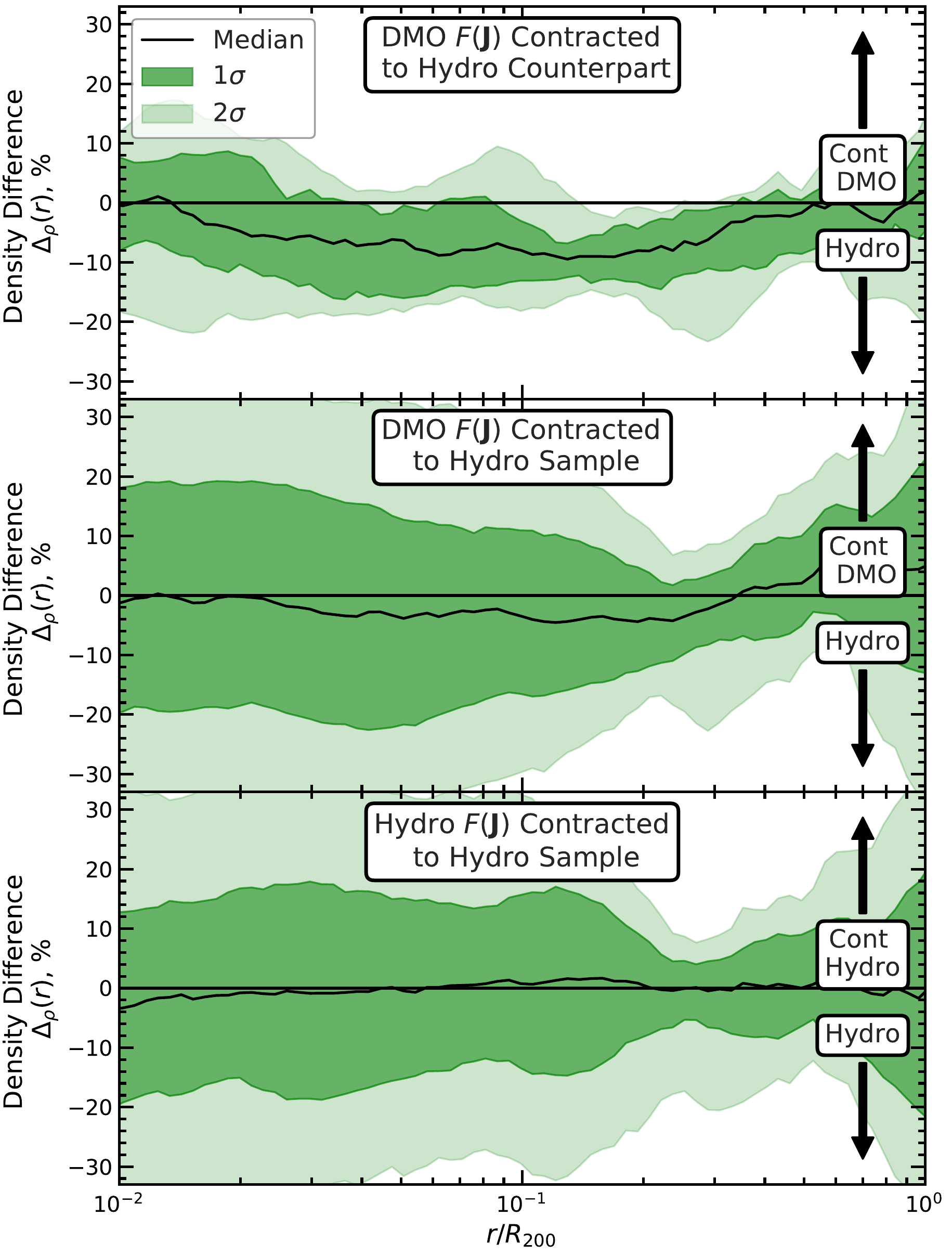}
	\vskip -.2cm
    \caption{ The difference in radial density profiles,
      $\Delta_{\rho}\left(r\right)$, between DM halos described by an
      action distribution, \FJrL{}, 
    adiabatically contracted according to a given baryonic profile
    (see main text for details) and the `true' DM halos in the
    \auriga{} hydrodynamical simulations. We show the results for
    relaxed halos only. The black line shows the median and the dark and
    light green regions indicate the 68\% and 95\% percentiles 
    respectively. In the top panel we compare contracted DMO halos
    with their Hydro counterparts, highlighting the effects  
    of unadibiatic differences in the action distributions between the DMO and Hydro simulations on the DM halo density profile. 
    In the middle panel we contract
    each DMO halo in turn to every other Hydro halo in the relaxed sample and compare the
    resulting density profiles, to additionally see the effects of halo variation. 
    In the bottom panel we contract each Hydro halo in turn to every
    other Hydro halo across the relaxed sample.
    This demonstrates the scatter expected when modelling an unknown contracted halo due to halo variation.
  }
    \label{fig:SampleDeltaDensity}
\end{figure}

We now apply the scheme of Section~\ref{Subsec:SelfConsistentHalo} to
model the DM halos in \auriga{}. The action distributions of the DM
halos, \FJrL{}, as found in Section~\ref{Subsec:OrbitalPhaseSpace},
are contracted to a fixed baryon potential,
$\Phi_{\textnormal{Baryon}}\left(r\right)$, taken from the
corresponding counterpart halo in the Hydro simulation.

First, we study if the \FJrL{} distribution measured in the DMO
simulation can be used to predict the DM distribution in the
counterpart Hydro halo. We illustrate this for the \Au5 halo in
Fig.~\ref{fig:HaloProp_r}, which shows the DM density as measured for
the Hydro halo (orange line) and the contracted DMO halo (blue
line). Although there is good overall agreement between the two, the
contracted halo density profile is slightly lower than the true one as
measured in the Hydro simulation. This systematic difference is
consistently seen in all the relaxed \auriga{} halos and is examined
further in the top panel of Fig.~\ref{fig:SampleDeltaDensity}, which
shows the fractional difference in density profiles between the
contracted DMO halo and the actual Hydro DM halo.  The contracted halo
systematically underpredicts the density profile by $\sim 8\%$ over
the radial range, $r\in[1/100,1/3]R_{200}$, while outside this range
the agreement is good. This results in $M_{200}$ masses for the
contracted halos that are $5 \pm 2\%$ lower than the true masses.  This
underprediction suggests a systematic, non-adiabatic, difference
between the Hydro and DMO action distributions, as we had already
encountered in Fig.~\ref{fig:OrbitalDist}.

To investigate the effects of halo-to-halo variations in action
distributions, we contract each of our relaxed DMO halos in turn
according to the baryonic distribution of each relaxed Hydro
halo. When doing so, we rescale the actions of the DMO halo to the total mass of each
target Hydro halo using the procedure described in
Section~\ref{SubSec:Scaling}, ensuring the final contracted halos have the
correct $M_{200}$. The fractional difference between the density
profiles of the contracted and 'true' halos are shown in the middle
panel of Fig.~\ref{fig:SampleDeltaDensity}. The variation in the DM
halos action distributions, \FJrL{}, produces a greater scatter in the
contracted density compared to when each halo is matched with its Hydro
counterpart. The scatter is largest in the inner third of the
halo beyond which the scatter is noticeably tighter before spreading
out again near the outskirts of the halo. This is likely due to the
variation in concentration, which mainly effects the inner regions of
the halo, $r\lesssim r_{\mathrm{s}}$. Alongside a greater scatter,
there is again an underprediction of the contracted density profile,
which is slightly reduced by fixing the mass of the contracted halos
to be equal to that of the Hydro halos.

We can overcome this systematic difference in the predicted density
profile by using the DFs measured in the Hydro halos instead of the
DMO halos, as we have done until now, as shown in the bottom panel of
Fig.~\ref{fig:SampleDeltaDensity}. The resulting contracted DM
profiles are unbiased but they have a rather large, ${\sim}15\%$,
halo-to-halo variation. This shows that the small systematic
differences we have seen in the actions between the DMO and Hydro
simulations (see Section~\ref{Subsec:OrbitalPhaseSpace}) have
measurable effects on the DM density profiles, and that to obtain
unbiased contracted DM halos we need to use the action distribution
measured in the Hydro simulations. Thus, to obtain an unbiased model
of the MW halo, we need to use Hydro derived DFs, and, because of
system-to-system variations in the DF, we can predict the MW halo
density profile only to 15\% accuracy.

We have studied in more detail the most important systematic
differences between the action DFs in the DMO and Hydro
simulations. The tests and the corresponding results are presented in
Appendix~\ref{Appendix:Adibiatic}. We have found that the small,
systematic difference in density profile seen in
Fig.~\ref{fig:SampleDeltaDensity} is predominantly driven by the
suppression of \Jr{} in the Hydro halos. 
In the Hydro simulations, some mechanism has caused the DM to lose radial energy in an unadibiatic way.
If the systematic decrease of radial action in the Hydro halos was driven by baryons through
either feedback or numerical baryon-DMO particle scattering effects we would perhaps expect to see the
strongest effects at the centre of the halo, where the baryon density
is highest. However, we see no evidence of a radially varying effect,
with the \Jr{} suppression being, on average, approximately the same at all radial distances from
the halo centre and at all angular momentum. Furthermore, the feedback driven cores found in some simulations of dwarf galaxies are formed by increasing the energy of the DM particles, not by reducing it.  We leave a more thorough investigation of
these non-adiabatic effects to future work.

\subsection{Local DM Properties in Auriga}\label{Subsec:AurigaLocalDM}

\begin{figure}
    \centering
	\includegraphics[width=\columnwidth]{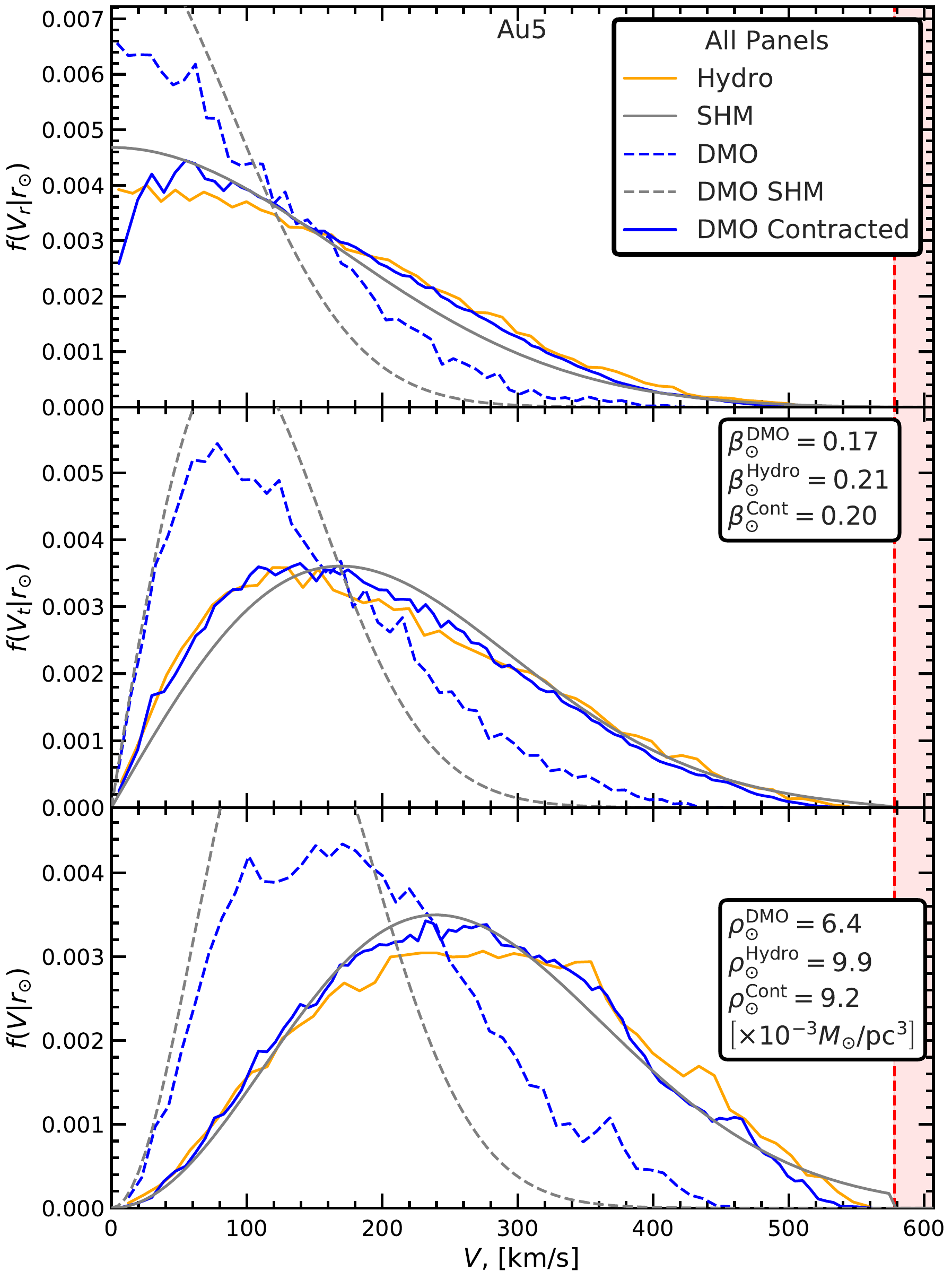}
	\vskip -.15cm
        \caption{The velocity distributions of DM particles at the
          Solar neighbourhood in the \auriga{} Halo~5, \Au5.  The
          solid orange line shows the distribution measured in the
          Hydro halo, the dashed blue line the corresponding quantity
          in the DMO case, and the solid blue line in the DMO halo
          contracted with our method to predict the Hydro
          quantities. In grey we show the predictions of the Standard 
          Halo Model (SHM), based on the assumption of an isotropic
          isothermal sphere. The top panel shows radial velocity,
          $v_{r}$, the middle panel tangential velocity, $v_{t}$, and
          the bottom panel total velocity, $v$.  The vertical red
          shaded region shows velocities larger than the escape
          velocity of the Hydro halo.  Estimates for the Solar
          neighbourhood DM density, $\rho_{\odot}$, and velocity
          anisotropy, $\beta_{\odot}$ are also given (see the two
          tables enclosed by a thick black line in the right-hand
          side of the centre and bottom panel).  }
    \label{fig:VelocityDist}
\end{figure}

As we discussed in the Introduction, a strength of the halo
contraction method presented here is that it can be used to predict
all DM halo properties, including the velocity distribution. This is
in contrast to most other methods
\citep[e.g.][]{Blumenthal1986ApJ...301...27B,Gnedin2004,CautunContract},
which apply only to the halo density profile. In this section we study
how the contraction method can predict dynamical properties of the DM
halo, in particular the DM velocity distribution in the Solar
neighbourhood, which is a crucial input into DM direct detection
experiments. 
In preparation for modelling the MW in Section~\ref{Sec:MW}, we first
study the velocity distribution function (VDF) of the relaxed DM
\auriga{} halos. To validate our methodology, we compare the
contracted DMO halos with their Hydro counterparts. Across our sample
of different size halos, we define an \auriga{} halo's `Solar
radius' as a set fraction of its $R_{200}$, 0.036$R_{200}$,
which was obtained by taking the following MW values:
$r_{\odot}=8~\textrm{kpc}$ and
$R_{200}^{\mathrm{MW}}\approx222~\mathrm{kpc}$ (from
\citetalias{CallinghamMass2019MNRAS.484.5453C}, corresponding to
$M_{200}^{\mathrm{MW}}=1.17\times10^{12}M_{\odot}$). 

We illustrate how well our contraction method recovers the DM velocity
distribution in the presence of a baryonic component by studying the
\Au5 halo. Compared to the DMO case, the Hydro halo has an enhanced
density and especially velocity dispersion at the Solar radius, as may
be seen in Fig.~\ref{fig:HaloProp_r}, at the radial position,
$r \sim 0.04 R_{200}$. The contracted DMO halo reproduces well the
Hydro halo, in particular, both the velocity dispersion as well as the
velocity anisotropy parameter, $\beta$. Thus, our contraction
technique reproduces local halo properties that are averaged over many
DM particles.

In Fig.~\ref{fig:VelocityDist} we show that the same technique also
reproduces the actual DM velocity distribution. For this, we calculate
the velocity distribution of all DM particles found within a radial
distance of $\pm1$ kpc around the Solar radius. As expected, the DM
particles in the Hydro case are characterised by higher velocities
than in the DMO case. The small irregularities in the distribution are
the result of the merger history of the halo. The action distribution
of the DMO halo can be used to predict the velocity distribution of
the contracted DMO halo. This is similar to the approach taken in
Sec. \ref{Subsec:SelfConsistentHalo}, where we modelled the density
profile. To obtain the VDF, we calculate the velocity components of
each \FrJrL{} orbit at the solar radius, and then sum over all
possible orbits, \FJrL{} (using a similar weighting to
Eqn.~\ref{Eq:PropertiesOrbits}).  The contracted DMO halo reproduces
well the velocity distribution of the Hydro halo, with most
differences between the two being stochastic in nature. The only large
difference is seen in the radial velocity, $v_{r}$, distribution (top
panel), where the contracted halo is systematically below the Hydro
case for $v_{r}\xrightarrow{}0$. This is due to the finite number of
orbits included in the reconstruction, with none being exactly at
apocentre, pericentre or on perfectly circular orbits at this
radius. This effect is small and can be reduced by including a greater
number of orbits in the reconstruction. 

The most popular approach in the field is to model the VDF using the
Standard Halo Model (SHM)
\citep[e.g.][]{EvansSHMPlus_2019PhRvD..99b3012E}. This is based on the
assumption of an isotropic isothermal sphere, and predicts a Gaussian
velocity distribution with velocity dispersion,
$\sigma=v_{\mathrm{circ}}/\sqrt{2}$, which is truncated at the escape
velocity, $v_{\mathrm{esc}}$.  The SHM predictions for the DMO and
Hydro simulations of \Au5 are shown in Fig.~\ref{fig:VelocityDist} as dashed
and solid grey curves, respectively. The SHM model provides a poor
description of the DMO velocity distribution, but performs much better
for the Hydro halo. However, we still find important differences
between the SHM predictions and the actual Hydro halos. In particular,
the sharp truncation of the SHM VDF at $v_{\mathrm{esc}}$ is more
abrupt than in the simulations and typically leads to an
overprediction of high velocity DM particles. Moreover, the SHM
assumes isotropic orbits whereas, in this halo and throughout our sample, we find a small, but non zero
anisotropy parameter at the Solar neighbourhood,
$\beta_{\odot}^{\mathrm{Hydro}}=0.21$. Thus, the isotropic SHM slightly
underpredicts the $v_{r}$ and overpredicts the $v_{t}$ distributions.

\section{Application to the Milky Way}
\label{Sec:MW} 

We can now apply our DM halo reconstruction method to infer the
structure of the DM halo of our own galaxy. To do this we need to
know: the action DF, \FJrL, of the MW halo; the MW baryon
distribution; and the total mass, $M_{200}^{\mathrm{MW}}$. The last
two quantities can be inferred from observations
\citep[e.g.][]{CautunContract,Wang2019:MW_mass_review}. For the
\FJrL{} distribution, we assume that the MW is a typical $\Lambda$CDM
halo and that its DF is similar to that of our relaxed \auriga{}
halos. By considering the range of different DFs for the MW, as
spanned by the \auriga{} halos, we quantify the extent to which the
unknown DM action distribution of our Galactic DM halo affects our
predictions.  Finally, as we saw in the previous section, there are
small systematic differences between the distributions of actions in
the DMO and Hydro simulations of MW-mass halos. Thus, to obtain
predictions that are as accurate as possible, we use the \FJrL{} DFs
measured in the Hydro simulations of the \auriga{} suite.

We adopt the MW baryon density profile advocated by
\citetalias{CautunContract}, which we model as a spherically symmetric
distribution. \citetalias{CautunContract} used parameterised density
profiles of a thick and thin stellar disc, a stellar bulge, a cold gas
ISM and an analytically contracted NFW DM halo. Using an MCMC fitting
procedure, these baryonic and DM components were fit to the latest MW
rotation curve data derived from \textit{Gaia} DR2 \citep{Eilers19Vcirc_2019ApJ...871..120E}
; the data used cover the
radial range 5 to 25 kpc. For the MW total mass, we adopt the value of
\citetalias{CallinghamMass2019MNRAS.484.5453C},
$M_{200}^{\mathrm{MW}}=1.17^{+0.21}_{-0.15}\times10^{12}M_{\odot}$.
This mass measurement was obtained by comparing the energy and angular
momentum of the classical MW satellites to the (E,L) distributions of
satellite galaxies in the \eagle{} simulation \citep{Schaye2015}.
This total mass determination is in very good agreement with other
measurements based on \textit{Gaia} DR2 (see Fig.~5 in
\citealt{Wang2019:MW_mass_review}), such as the ones based on escape
velocity
\citep{Deason19Vesc_2019MNRAS.485.3514D,Grand_DMVescMass_2019MNRAS.487L..72G},
globular cluster dynamics
\citep{PostiHelmi2018arXiv180501408P,WatkinsMWGC2018arXiv180411348W}
and rotation curve modelling \citepalias{CautunContract}.

\begin{figure}
	\includegraphics[width=\columnwidth]{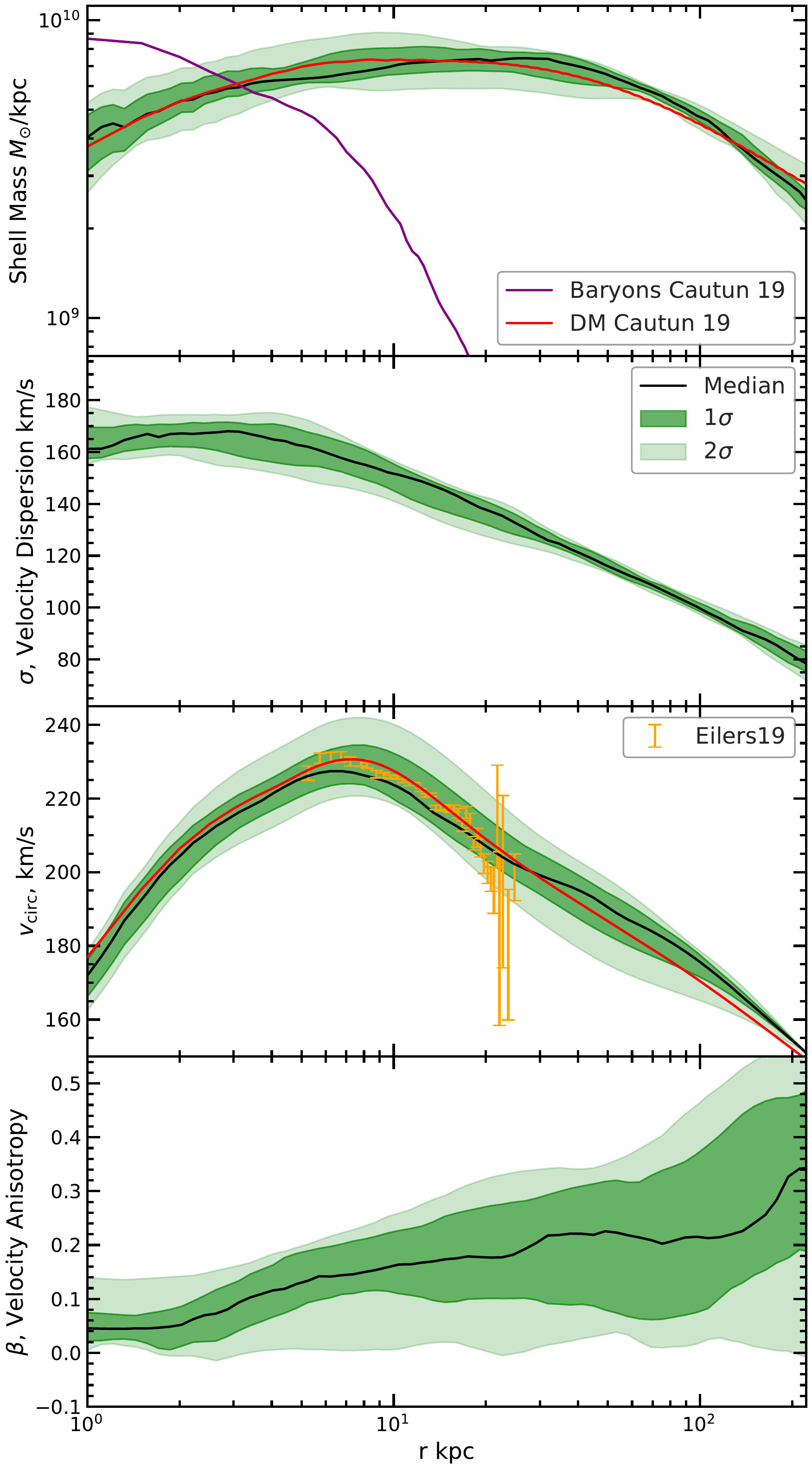}
	\vskip -.15cm
	\caption{From top to bottom: the MW's density, velocity
          dispersion, circular velocity, and velocity anisotropy
          radial profiles predicted by our halo contraction
          method. The DM halos are contracted assuming the
          \citetalias{CautunContract} MW baryonic model and the action
          distributions, \FJrL{}, from 17 relaxed halos in the
          \auriga{} Hydro simulations. The black line shows the median
          prediction of our method, while the dark and light shaded
          regions show the 68 and 95 percentiles arising from
          halo-to-halo variation in \FJrL{}.  The top panel also shows
          the \citetalias{CautunContract} baryonic profile (purple)
          and their best fitting DM profile (red) ; the third panel
          shows in yellow the \protect\cite{Eilers19Vcirc_2019ApJ...871..120E} $V_{\rm{circ}}$
          data and in red the \citetalias{CautunContract} rotation
          curve for their best fitting MW model.}
	\label{fig:MWProp_r}
\end{figure}

Our inferred properties of the MW DM halo are shown in
Fig.~\ref{fig:MWProp_r}. In the top panel we see that the median of
the contracted density profile closely matches that of
\citetalias{CautunContract}, although some differences are
present. This is to be expected since the \citeauthor{CautunContract}
results corresponds to a DM halo that, before baryon contraction, had
a concentration of 8.2, while the 17 \auriga{} halos studied here have
a wide range of concentrations (see
Fig.~\ref{fig:AurigaConc}). Nonetheless, the
\citeauthor{CautunContract} result lies well within the 68 percentile
scatter of our predictions, indicating good overall agreement.  Not
knowing the exact \FJrL{} distribution of the MW halo results in a
$\sim 14\%$ scatter (68 percentile range) in the predicted density
profile of the contracted halo, in good agreement with our \auriga{}
results. Our model predicts that the DM velocity dispersion is roughly
constant at around $160\kms$ in the inner region of our Galaxy, and
then decreases rapidly towards the halo outskirts (second panel in
Fig.~\ref{fig:MWProp_r}).

In the third panel of Fig.~\ref{fig:MWProp_r}, we compare the circular
velocity curve predicted by our model with the actual estimate for the
MW as determined by \cite{Eilers19Vcirc_2019ApJ...871..120E}. We do
not fit our model to these data, so the good agreement with
observations indicates that our model is making sensible
predictions. To compare against the data, we add the circular velocity
curves from both the baryons and the halo. The latter is modelled as a
spherically symmetric distribution but for the baryons we need to take
into account that their distribution is highly flattened, i.e. most
stars and gas are found in a disc, and that the
\citeauthor{Eilers19Vcirc_2019ApJ...871..120E} rotation curve is
measured in the plane of this disc.  In the plane of the disc, the
true axisymmetric profile gives a ${\sim}$10\% greater contribution to
the circular velocity than the spherical profile that we use when
modelling the contraction of the DM halo.

The distribution of circular velocity curves across our contracted DM
halos are in good agreement with both the
\citeauthor{Eilers19Vcirc_2019ApJ...871..120E} data and the
\citetalias{CautunContract} best fitting model. However, we see
variation in the curves when using different action DFs. This is to be
expected since the MW represents one possible realisation of
\FJrL{}. It is worth stressing that the median result is not
necessarily the `best' model for the MW DM halo, as the MW is unlikely
to reside in a typical $\Lambda$CDM halo. Instead, the point to emphasise
is that we would expect the MW to lie within the range of our halo
sample, i.e. within the scatter, which it clearly does.

While the \citeauthor{Eilers19Vcirc_2019ApJ...871..120E} circular velocity curve data lie comfortably within
the $1\sigma$ range of our distribution of contracted halos, the
individual halo curves are poor fits. This shortcoming could be
overcome by using the observations to find out which \FJrL{}
distribution best describes the MW data. This can be achieved with a
MCMC approach in which we sample different action DFs and concurrently
constrain the MW baryonic distribution \citepalias[e.g. similar to the
approach of][]{CautunContract}. It is important to marginalise over
the MW baryonic distributions, since these are uncertain and, as
\citetalias{CautunContract} have shown, there is a degeneracy between
the baryon content and the DM halo structure when modelling the MW
rotation curve. This approach is beyond the scope of this paper and we
leave it for future work.

\subsection{MW Local DM Distribution}\label{SubSec:MW_DM}

\begin{figure}
    \centering
	\includegraphics[width=\columnwidth]{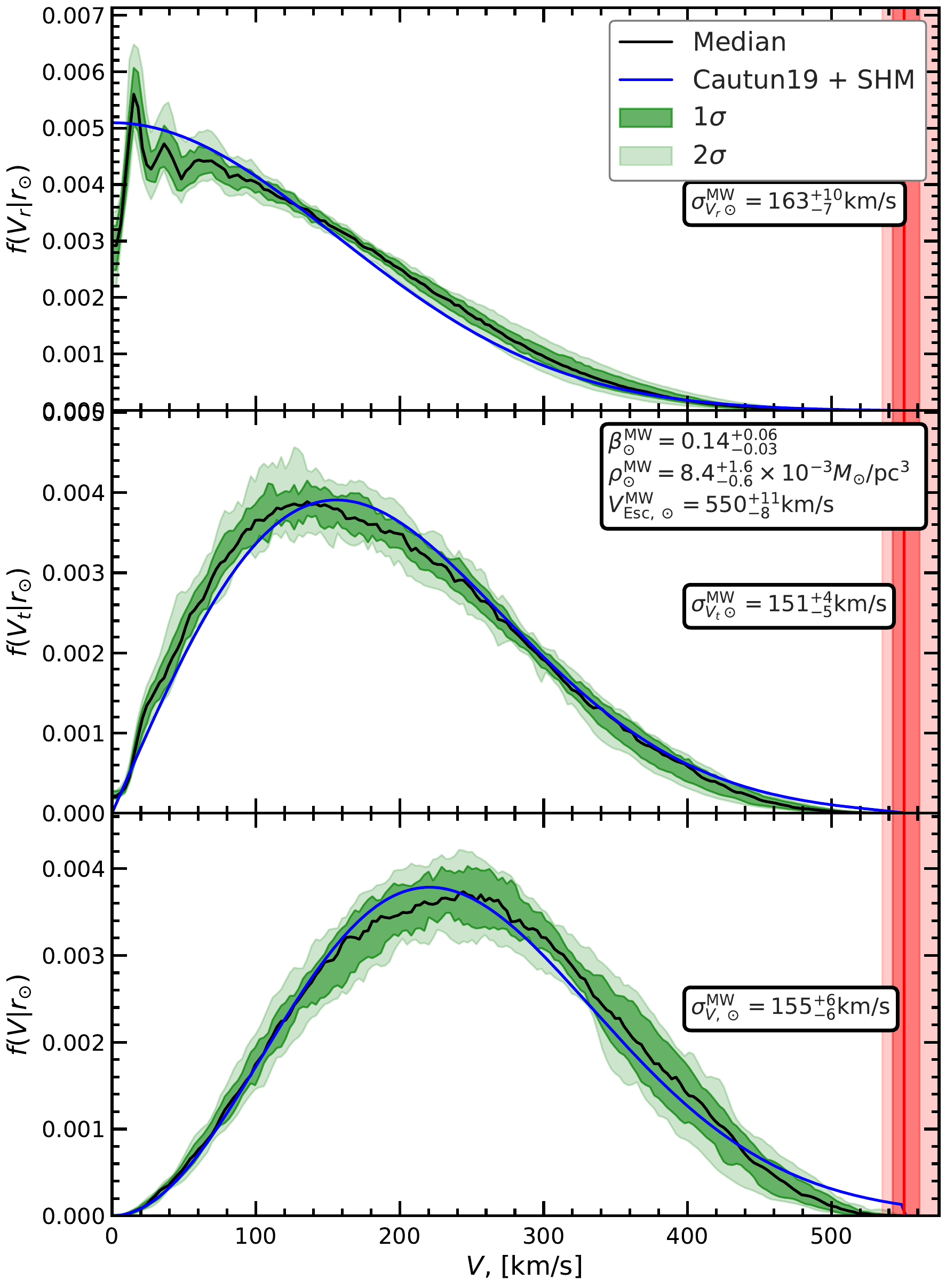}
	\vskip -.15cm
        \caption{The DM velocity distribution at the Solar radius,
          $r_{\odot}=8 \mathrm{kpc}$, as predicted by our halo
          contraction model.  The radial, tangential, and total
          velocity distributions are shown in the top, middle and
          bottom panels, respectively. The median is indicated with a
          solid black line and the 68 and 95 percentiles are shown in
          shaded green. The blue curve illustrates the velocity
          distributions given by the Standard Halo Model (SHM) using
          the \citetalias{CautunContract} MW mass model. We also give
          several DM properties at the Solar radius (see text inserts
          in the panels) as predicted from our model: the local DM density, $\rho_{\odot}$; the
          components of the velocity dispersion, $\sigma_{\odot}$; 
          the velocity anisotropy, $\beta_{\odot}$; and the escape
          velocity, $V_{\mathrm{Esc},\odot}^{\mathrm{MW}}$ (whose
          value is also shown as the red shaded region).}
    \label{fig:MWVelocityDist}
\end{figure}

\begin{table}
	\centering
	\caption{A list of MW properties in the Solar neighbourhood
          inferred from our DM halo contraction model. The third
          column gives the corresponding values from
          \citetalias{CautunContract}. Note the velocity dispersion's and anisotropy of this column are not found directly in \citetalias{CautunContract}. Instead these values, denoted by *, are
          calculated by applying the SHM to the
          \citetalias{CautunContract} MW mass distribution.
	}
	\label{Tab:LocalDM}
    \setlength\extrarowheight{5pt}
\begin{tabularx}{1\columnwidth}{l l c r}

\hline
\hline

Property &  This work  & Cautun19 + SHM*      & Units \tabularnewline[.1cm]

\hline
$\rho_{\odot}$ & $8.4^{+1.6}_{-0.6}$ & 9.2 & $10^{-3} M_{\odot}/ \mathrm{pc}^{3}$ \tabularnewline

                     &$0.32^{+0.06}_{-0.02}$& $0.34^{+0.02}_{-0.02}$ & $\mathrm{GeV/cm}^3$
                                                                      \tabularnewline

$\sigma_{V,\odot}$ & $155^{+6}_{-6}$ & 156*   & $\mathrm{km/s}$ \tabularnewline

$\sigma_{V_{r},\odot}$ & $163^{+10}_{-7}$ & 156*   & $\mathrm{km/s}$ \tabularnewline

$\sigma_{V_{t},\odot}$ & $151^{+4}_{-5}$ & 156*  & $\mathrm{km/s}$ \tabularnewline

$\beta_{\odot}$ & $0.14^{+0.05}_{-0.03}$ & 0 (isotropic)*  & -- \tabularnewline

$V_{\mathrm{Circ},\odot}$ & $226^{+8}_{-3}$ & 230  & $\mathrm{km/s}$ \tabularnewline 

$V_{\mathrm{Esc},\odot}$ & $550^{+11}_{-8}$ & 549  & $\mathrm{km/s}$ \tabularnewline 

$M_{200}^{\mathrm{Total}}$ & $1.17$ & $1.12$ & $10^{12}M_{\odot}$ \tabularnewline 

$M_{200}^{\mathrm{DM}}$ & $1.04$ &1.00 &  $10^{12}M_{\odot}$ \tabularnewline 

$M_{200}^{\mathrm{Baryons}}$ & $0.13$ &0.13 & $10^{12}M_{\odot}$ \tabularnewline[.1cm]
\hline

\end{tabularx}
\end{table}

Having inferred the likely structure of the MW DM halo by applying the
results of Section~\ref{Sec:HaloConstruction} based on analysis of 17 \auriga{} galactic halos, we now investigate the implications for the
key astrophysical inputs to direct DM detection experiments: the
density and velocity distribution of the DM in our own solar
neighbourhood.

From the DM density profile shown in the top panel of
Fig.~\ref{fig:MWProp_r} we find that our models predict
$\rho_{\odot}=8.4^{+1.6}_{-0.6}\times 10^{-3} M_{\odot}/
\mathrm{pc}^{3}$ (equivalently
$\rho_{\odot}=0.32^{+0.06}_{-0.02}/
\mathrm{GeV}/\mathrm{cm}^{3}$). This values are in good agreement with
previous estimates \citep[see the compilation
by][]{Read2014}. The somewhat large uncertainties in our estimate of
$\rho_{\odot}$ could be significantly reduced if we were to restrict our analysis to those DFs
that best fit the MW rotation curve, or individually fit the MW baryon distribution for each DM halo as discussed at the end of the previous section.

In Fig.~\ref{fig:MWVelocityDist} we highlight the DM velocity distributions at the Solar position predicted by our MW
  models. These were derived using the method described in
  Section~\ref{Subsec:AurigaLocalDM} where, for each model, we sum the
  orbits of all DM particles to find the distribution of radial,
  tangential and total velocity components. The resulting VDFs have
  very similar forms to those previously discussed for the \Au5 halo (see
  Fig.~\ref{fig:VelocityDist}) and many of the conclusions reached for
  that example apply here too. In particular, we predict a small radial
  bias in the velocity anisotropy,
  $\beta_{\odot}=0.14^{+0.07}_{-0.03}$, with the radial and tangential
  velocity dispersions being
  $\sigma_{V_{r},\odot}^{\mathrm{MW}}=163^{+10}_{-7}\mathrm{km/s}$ and
  $\sigma_{V_{t},\odot}^{\mathrm{MW}}=151^{+4}_{-5}\mathrm{km/s}$.
  These and other values are summarised in Table~\ref{Tab:LocalDM},
  where we also compare our results to those from the recent MW mass
  model of \citetalias{CautunContract}. In this table the velocity dispersion's and anisotropy given for \citeauthor{CautunContract}  are the results of applying the SHM
  with the parameters inferred from the \citeauthor{CautunContract} MW
  mass model. See Sec. \ref{Subsec:AurigaLocalDM} for further details and discussion on the SHM.

  The SHM is in good overall agreement with our inferred velocity
  distribution, although we find large fractional deviations in the
  high velocity tail of the distribution, the region to which DM
  direct detection experiments are most sensitive
  \citep{Bozorgnia2019b}. The SHM model assumes an isotropic velocity
  distribution, at odds with the value of $\beta_{\odot}\sim0.14$ in
  our model. As a result, the SHM does not perform as well when
  compared against the radial and tangential velocity distribution of
  our model. We find a local speed escape velocity
  $550^{+11}_{-8}\mathrm{km/s}$, which is consistent with the recent
  \textit{Gaia} DR2 measurements of
  \citet{Deason19Vesc_2019MNRAS.485.3514D} and
  \citet{Grand_DMVescMass_2019MNRAS.487L..72G}.

\section{Conclusions}
\label{Sec:Conclusion} 

We have used the \auriga{} suite of hydrodynamical simulations of
Milky Way (MW) analogues to investigate the orbital distribution of DM
particles in MW-mass halos and to study how this distribution changes
when including baryonic physics in the simulations. We have
characterised the DM halos in terms of the distribution of spherical
actions: radial action, \Jr{}, and angular momentum, $L$. We have
studied these action DFs for all our relaxed halos and have described
how the actions can be used to (re)construct the density and velocity
distribution of the simulated DM halos. This can be achieved using an
iterative method that, starting from a fixed baryonic distribution and
an initial guess for the gravitational potential, constructs a DM halo
density profile. At each step in the iteration the potential is
updated from the DM mass profile obtained in the previous step until
convergence is achieved.

The actions \Jr{} and $L$ are useful quantities for describing DM
halos since they are conserved during adiabatic changes (i.e. on long
timescales) in the gravitational potential. Many galaxy formation
processes, although not all, are thought to be adiabatic and this
suggests that halos in dark matter only (DMO) and Hydro simulations
should have similar \FJrL{} distribution functions. This idea
motivated us to investigate if indeed the action DF is conserved in
the \auriga{} suite between the DMO simulations and the simulations
that include galaxy formation physics. We have found good agreement
between the actions in the DMO and Hydro halos, with differences at
the $5-10\%$ level. Most of these differences are due to statistical
fluctuations; however, we also find systematic variations, with \Jr{}
being lower in the Hydro halos.  This difference in radial action
leads to an $\sim 8\%$ underprediction of the DM density profile when
adiabatic contraction of a DMO halo is assumed.  The \Jr{} systematic
difference is the same at all radii, suggesting that it is unlikely to
be caused by effects associated with baryonic feedback which would mainly
affect the central region of a halo.

If we know the \FJrL{} actions of a halo in a DMO simulation, we can
predict the density and velocity profile of its counterpart in the
hydrodynamical simulation with a precision of ${\sim}5\%$ (not
withstanding the systematic effects discussed above). Most of the
scatter is due to stochastic effects as well as to small deviations
from the steady state assumption implicit in our method. This
object-to-object scatter is a factor of two lower than for other
methods, such as that of \citetalias{CautunContract}.  However, if we
do not know the exact \FJrL{} distribution, we recover the density
profiles only with ${\sim}15\%$ precision, with the major limitation
being the halo-to-halo scatter in the action distributions.

We have illustrated the contraction of a DM halo in the presence of
baryons by decomposing the halo into individual orbits of DM
particles. The deeper potential in the Hydro case leads to a
contraction, i.e. an inward shift, of the orbits. For a fixed orbital
angular momentum, circular orbits contract the most while highly
elliptical orbits contract the least. The DM halo is specified by the
sum of all orbits as given by the \FJrL{} distribution. This property
can be used to determine both the density and velocity distribution
profiles of a halo.

We have applied our DM halo construction method to the halo of the
MW. Starting from the \FJrL{} distribution of relaxed \auriga{} Hydro
halos, in combination with the \citetalias{CautunContract} stellar and
gas model of the MW and the value of the MW mass of
\citetalias{CallinghamMass2019MNRAS.484.5453C}, we have predicted the
density and velocity distribution of our galaxy's DM halo. This
resulted in 17 models for the Galactic DM halo, which span possible DM
distributions given the MW's baryonic component. We find good
consistency between our inferred DM halo density and that inferred by
\citeauthor{CautunContract}, and between the circular velocity curve
predicted by our models and the one measured from \textit{Gaia} DR2
data \citep{Eilers19Vcirc_2019ApJ...871..120E}. The consistency with
the \citeauthor{CautunContract} results provides an independent check
that their DM halo contraction model gives a good description of the
Galactic DM distribution.

A major advantage of our halo (re)construction method is that it can
predict the velocity distribution of DM particles. We have tested this
aspect of our method by comparing directly against measurements of the
\auriga{} halos and found very good agreement. In particular, our
method does better than the Standard Halo Model (SHM) at reproducing
the high tail of the velocity distribution, a key input into direct DM
detection experiments. We have applied the same analysis to the MW to
predict the distribution of DM particle velocities and their
components in the solar neighbourhood. Our results are in good
agreement with the literature
\citep[e.g.][]{EvansSHMPlus_2019PhRvD..99b3012E,Bozorgnia2019b}, and
predict that the DM particles have a preference for radial orbits,
with $\beta_{\odot}=0.14^{+0.07}_{-0.03}$, and that the SHM
overpredicts the high velocity tail of the velocity
distribution. Furthermore, by using multiple action distributions, we
have characterised the halo-to-halo scatter in the velocity
distribution, which is important for understanding how robust are the
constraints inferred from direct DM detection experiments.

Our work leaves open an important question: which baryon processes are
responsible for the systematic difference in the action distribution
between the DMO and the Hydro halos? While such effects are small,
about a few percent, they produce a measurable effect on the density
profile and velocity distribution. To overcame this systematic when
modelling the MW, we have used the \FJrL{} distribution measured
directly in the Hydro simulations. It remains to be seen if the same
systematic deviations between DMO and Hydro halos are present in other
simulations and if the size of the effect varies between the various
subgrid galaxy formation models implemented in different simulations.

In this work, when making predictions specifically for the MW, we
employ a range of possible action distribution functions of a MW-mass halo
as predicted by the \auriga{} project. However, given
the observations, e.g. the MW rotation curve, some distributions are
more likely than others. This raises the question of which is the best
fitting \FJrL{} distribution for the MW, which we leave for future
work. To address this will require  modelling the
still uncertain MW baryon mass distribution self-consistently alongside the DM distribution, since this is degenerate
 when predicting the inner
($\lesssim50~\rm{kpc}$) rotation curve \citepalias[for details
see][]{CautunContract}. Such a study is very worthwhile and timely,
especially given the wealth of Galactic data available in the current
and future \textit{Gaia} data releases. 

The method we have presented here provides a very comprehensive tool
for modelling DM halos in the presence of baryons and, furthermore, it
can easily account for cosmological halo-to-halo variations in halo
properties.  In the age of precision MW astronomy it is no longer
possible to neglect the contraction of the Galactic DM halo or the
diversity of DM distributions that form a halo. Our method provides an
elegant and robust approach to incorporate these effects.


\section*{Acknowledgements}

TC, MC and CSF were supported by the Science and Technology Facilities
Council (STFC) [grant number ST/F001166/1, ST/I00162X/1,ST/P000541/1]
and by the ERC Advanced Investigator grant, DMIDAS [GA 786910].  MC
also acknowledges support by the EU Horizon 2020 research and
innovation programme under a Marie Sk{\l}odowska-Curie grant agreement
794474 (DancingGalaxies).  AD is supported by a Royal Society
University Research Fellowship. CSF acknowledges European Research
Council (ERC) Advanced Investigator grant DMIDAS (GA 786910). FM acknowledges support through
the Program ‘Rita Levi Montalcini’ of the Italian MIUR. This
work used the DiRAC Data Centric system at Durham University, operated
by ICC on behalf of the STFC DiRAC HPC Facility
(www.dirac.ac.uk). This equipment was funded by BIS National
E-infrastructure capital grant ST/K00042X/1, STFC capital grant
ST/H008519/1, and STFC DiRAC Operations grant ST/K003267/1 and Durham
University. DiRAC is part of the National E-Infrastructure.




\bibliographystyle{mnras}
\bibliography{ContractionPaper.bib} 



\appendix

\section{Radial distribution of orbits}
\label{Appendix:Orbits}

Here we describe how to construct the probability distribution that a
particle on a orbit defined in terms of $(\Jr,L)$ is found at radial
distance, $r$. We denote this radial probability distribution as
$F\left(r|\Jr,L\right)$. For simplicity, in the following we will work
with the $(E,L)$ actions (and thus calculate $F\left(r|E,L\right)$),
which, given a gravitational potential, can be uniquely mapped to 
$(\Jr,L)$ space and viceversa (see main text for details).

Consider an orbit defined by $(E,L)$ in the potential $\Phi\left( r
\right)$. The velocity components at $r$ are defined as: 
\begin{equation}
    \begin{aligned}
    v|_{r}=&\sqrt{2\left(E-\Phi\left(r\right)\right)}  \\
    v_{t}|_{r}=& L/r  \\
    v_{r}|_{r}=&\sqrt{v|_{r}^{2}-v_{t}|_{r}^{2}}, 
    \end{aligned}{}
\end{equation}
where for the radial velocity component we only consider its absolute
value. A tracer on that orbit could have either negative or positive
$v_{r}$ depending on whether it is approaching or receding from the halo
centre. The two points where $v_{r}=0$ correspond to the peri- and
apocentre of the orbit, $r_{\mathrm{min}}$ and $r_{\mathrm{max}}$,
with particles on the orbit spanning the radial range, 
$r_{\mathrm{min}}<r<r_{\mathrm{max}}$.

As described in the main text, the radial distribution of an orbit,
either $F\left(r|E,L\right)$ or $\FrJrL$, is defined as the proportion
of time an orbit spends at radial distance, $r$, normalised to
unity. To calculate this, we first consider the amount of time, \dt{},
taken by a test particle to travel from $r\xrightarrow{} r+dr$. By
Taylor expansion, we have
\begin{equation}
    r+\dr=r+v_{r}\dt+\frac{1}{2}\ddot{r}{\dt}^{2}+o\left({\dt}^{3}\right)
    \; ,
\end{equation}
where $\ddot{r}$ denotes the radial acceleration, i.e. the second
derivative of $r$ with respect to time.  By neglecting $dt^3$ and
higher order terms, we can solve for \dt{} to obtain
\begin{equation}
    \dt=\frac{-v_{r}+\sqrt{v_{r}^{2}+2\ddot{r}\dr}}{\ddot{r}}
    \;.
\end{equation}

Away from the endpoints, $v_{r}^{2}\gg2\ddot{r}\dr$ for small
\dr. Then $\dt \approx \dr/v_{r}$, that is the time spent at $r$ is
inversely proportional to the radial velocity component, as
expected. As the test particle approaches the endpoints,
$v_{r}\xrightarrow{}0$ and the radial acceleration terms can no longer
be neglected.  Then, the fraction of time spent at $r$, i.e. the radial
distribution $F\left(r|E,L\right)$, can be written as
\begin{equation}
    F\left(r|E,L\right)\mathrm{d}r=\frac{2}{T|_{E,L}}\dt|_{r}
    \; ,
\end{equation}
where the factor of $2$ accounts for the fact that a particle is found
at the same $r$ value twice along its orbit, i.e. once when
approaching and once when receding from the halo centre. The
normalisation factor, $T|_{E,L}$, is the radial time period, which is
given by
\begin{equation}
    T|_{E,L}=2\int_{r_{\mathrm{min}}}^{r_{\mathrm{max}}}\mathrm{d}t|_{r}\mathrm{dr}. 
\end{equation}

To calculate \FrEL\ we use a radial grid with 1500 cells defined in the
range $\left[0,3R_{200} \right]$; this corresponds to a grid spacing,
$\dr = R_{200}/500 \sim 0.5{\rm kpc}$.  Special treatment is required at the endpoints of
  the orbit where better spatial resolution is needed to track the orbit
  properly. The radial distribution and properties around
  $1\mathrm{kpc}$ of the end points of each orbit are then
  recalculated at a higher radial resolution of
  $\dr^{*} = 5\mathrm{pc}$.
  
Averaged radial properties, such as the velocity dispersion or the
velocity components, can be evaluated at a given radius using \FrJrL{}
as a weight. Any general orbital property depending on radius,
$X\left(r\right)|_{J_{r},L}$, can be calculated as

\begin{equation}
    \label{Eq:PropertiesOrbits}
    \left<X\right>\left(r\right)=\frac{1}{\rho\left(r\right)}\frac{M}{4\pi r^{2}}\iint X\left(r\right)|_{J_{r},L} \FrJrL dJ_{r}dL
    \;.
\end{equation}

\section{Systematic Differences in Action between DMO and Hydro}\label{Appendix:Adibiatic}

\begin{figure}
	\includegraphics[width=\columnwidth]{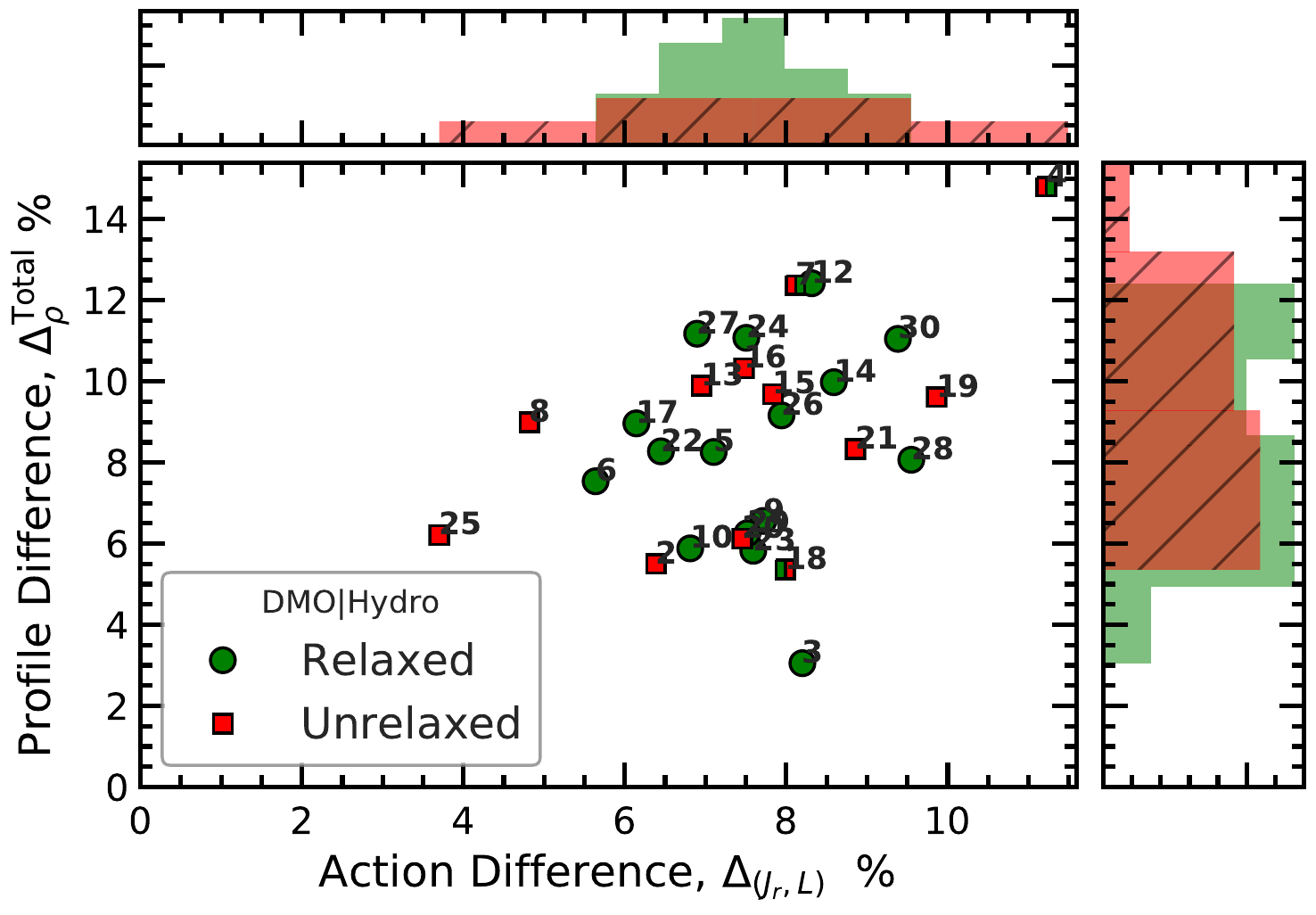}
	\vskip -.1cm
        \caption{An exploration of the extent to which galaxy
          formation in the \auriga{} suite is an adiabatic
          process. The $x$-axis show the difference in the action
          distributions, $\Da$, of DM halos between the DMO and
          corresponding Hydro simulations. An adiabatic process would
          conserve the action, i.e. $\Da=0$. The $y$-axis shows the
          integrated difference, $\Dr^{\rm total}$, in the DM density
          profiles between the contracted DMO halo and the Hydro
          halo. Each symbol represents one \auriga{} system and the
          green or red colours indicate if the halo is relaxed or
          unrelaxed. (See text for definitions and further details.)
          In the relaxed sample, the $\Da$ and $\Dr^{\rm total}$
          quantities show only a slight correlation (0.16), suggesting
          a complex relationship between differences in action
          distributions and differences in the final contacted
          profile.  }
    \label{fig:AdiScatter}
\end{figure}

Differences between DM halos, such as in the $\rho\left(r\right)$, $\sigma\left(r\right)$ and
$\beta\left(r\right)$ profiles, can be attributed to differences in
their action distributions, \FJrL{}. It is natural to expect that the
greater the action difference, \Da{}, between our DMO and Hydro halos,
the greater the difference in the contracted DM density profile.  We
explore this correlation in Fig.~\ref{fig:AdiScatter}, which shows the
integrated difference in the density profiles, $\Dr^{\rm total}$,
between the contracted DMO halo and the Hydro halo as a function of
the difference in the action, \Da{}, between the two halos. In the
relaxed halo sample, the $\Dr^{\rm total}$ and \Da{} quantities are
characterised by a small correlation of only 0.16.
This suggests a complex relationship between action distributions and the physical halo.  The relaxed sample
has consistent differences of $\Dr^{\rm total} \sim 8\% \pm 3\%$,
while the unrelaxed sample has a a wider scatter and a higher median
$\Dr^{\rm total} \sim 10_{-3}^{+10}\%$ (a histogram of the results may
be seen in the side panel of Fig.~\ref{fig:AdiScatter}).

\begin{figure}
    \centering
	\includegraphics[width=0.95\columnwidth]{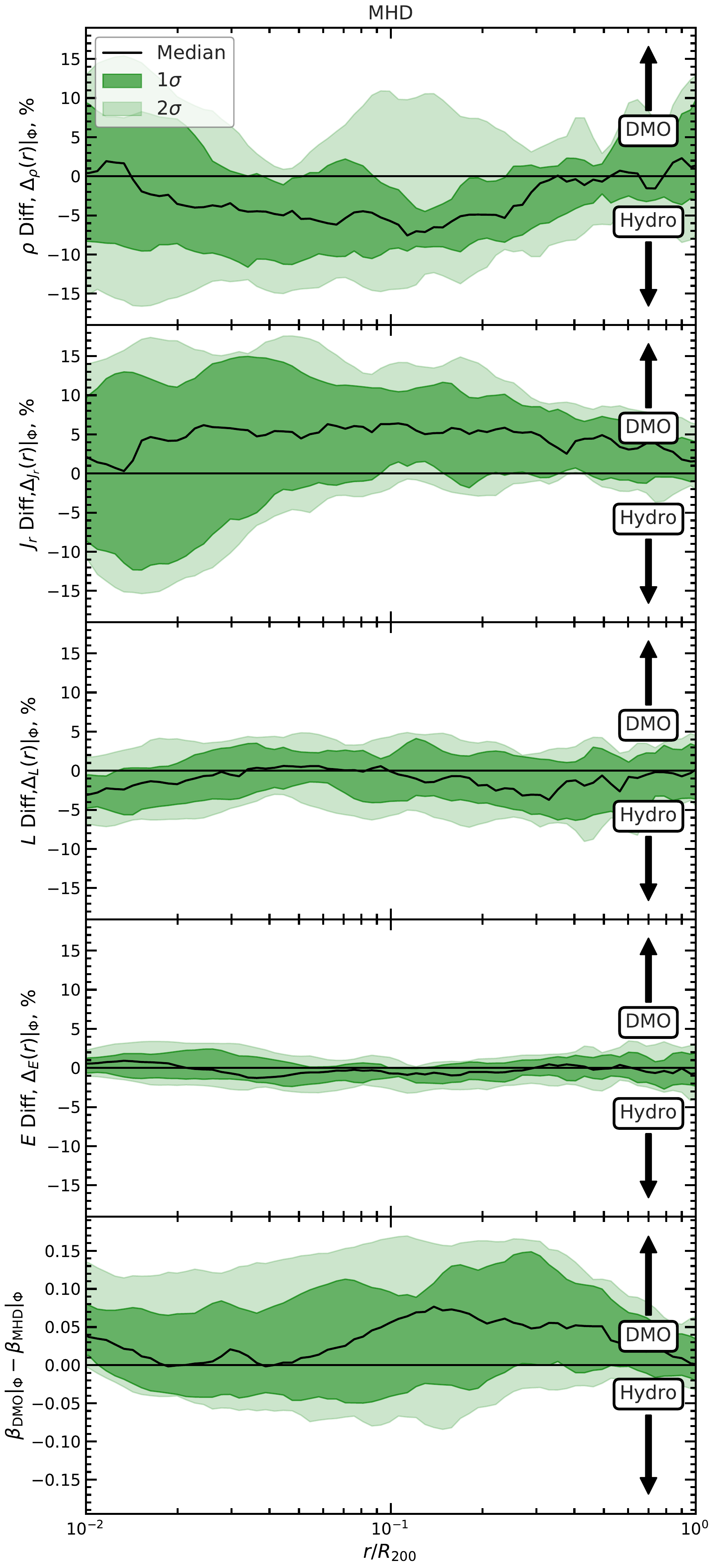}
	\vskip -.1cm
	\caption{The fractional differences in selected halo
          properties as a function of radial distance.  We plot the
          difference between halo quantities calculated using the
          \FJrL{} distribution measured in the DMO and in the
          counterpart Hydro simulation. When reconstructing the halo
          properties we use \textit{the same fixed potential},
          $\Phi_{\mathrm{Hydro}}$, measured in the Hydro simulation; in this
          way any difference in the plotted quantities are due to
          variations in the action distribution between the DMO and
          Hydro halo, and not to changes in the potential. We show,
          from top to bottom, the differences in the radial profile
          of: density, average radial action, average angular momentum,
        average energy, and average velocity anisotropy. The black
        line gives the median for our sample of relaxed \auriga{}
        halos, and the dark and light green regions the 68 and
        95 percentiles, respectively. The DMO density profile, $\rho$
        (top panel), is systematically lower than in the Hydro
        counterpart, driven by a systematic suppression of radial
        action in the Hydro halo at every radius (second panel). 
	}
	\label{fig:Diffr}
\end{figure}

To better understand the effect of systematic differences in the
\FJrL{} distribution between the DMO and Hydro simulations, we proceed
to compare in Fig.~\ref{fig:Diffr} the radial profiles of several halo
properties. in the main text, when constructing the DM density profile given
a \FJrL{} distribution, we find the self-consistent gravitational
potential given the action distribution. However, differences in
actions can lead to differences in potentials that would further
enhance differences in DM halo properties. To control for changes in
potential, the results in Fig.~\ref{fig:Diffr} are obtained by
constructing the DM halos using the same fixed potential,
$\Phi_{\mathrm{Hydro}}$, measured in the Hydro simulation. 
This allows a direct comparison of the orbital structure in physical
space, providing insight into the dependence of the differences in
density profile on the differences in action distributions. The
potential mechanisms behind non-adiabatic effects can also be explored
through the radial dependence of the action differences.

In Fig.~\ref{fig:Diffr} we consider the fractional differences in the
density and average actions as a function of radius. In the top panel
we  see a $\sim5\%$ underprediction of the DM density when using
 actions of the DMO halo compared to the Hydro. 
The slightly changed potential generated with this density profile causes the density difference to grow with the itteration to  $\Dr^{\rm total} \sim 8\%$ in the final self consistent profile. 
 For $L$, we find very small
systematic differences, but nonetheless the Hydro simulations tend to
have slightly higher $L$ values in the very inner regions and for
$r\sim 0.3 R_{200}$. In contrast, the energy distribution is
characterised only by small stochastic differences.

 The \Jr{} in the DMO halos is
systematically higher at all radii away from the very centre
$r \gtrsim 0.1 R_{200}$ (second panel). 
For a single
orbit, increasing \Jr{} causes the median position of an orbit
$\left<r\right>|_{\textrm{Orbit}}$ to move radially outward, and mass
to move from the radial centre of the orbit to its endpoints, as seen
in Fig.~\ref{fig:OrbitRadDist}. This effect across all orbits seems to drive
the difference in density profile (top panel): the density is higher
in the Hydro halos at intermediate radii, but the density is higher in
the DMO halos at the centre and near $R_{200}$. The higher radial
action gives more radial orbits in the DMO case, increasing
$\beta_{\mathrm{DMO}}$ (see bottom panel of Fig. \ref{fig:Diffr}).

For a discussion of how the results shown in Fig.~\ref{fig:Diffr} can
be used to understand the effects driving the non-adiabatic change in
DM actions between the \auriga{} DMO and Hydro simulations, we refer
the reader to the last paragraph of Section~\ref{Subsec:ContractAuriga}.



\bsp	
\label{lastpage}
\end{document}